\documentclass[sigconf]{aamas} 

\usepackage{balance} 

\usepackage{braket}
\usepackage{amsmath}
\usepackage{amsthm}
\newtheorem{assumption}{Assumption}
\newtheorem{problem}{Problem}
\usepackage{algorithm}
\usepackage{algpseudocode}
\usepackage{multirow}
\usepackage{etoolbox}
\usepackage{subcaption}   
\usepackage{graphicx}
\usepackage{svg}
\usepackage{hyperref}
\hypersetup{
    colorlinks=true,
    urlcolor=blue,  
    linkcolor=black,
    citecolor=black
}

\newboolean{reviewtextflag}
\setboolean{reviewtextflag}{true}

\newcommand{\reviewColor}[1]{%
  \ifthenelse{\boolean{reviewtextflag}}{\textcolor{blue}{#1}}{#1}%
}

\DeclareMathOperator*{\argmin}{arg\,min}

\usepackage{algpseudocode}

\algnewcommand\algorithmicforeach{\textbf{for each}}
\algdef{S}[FOR]{ForEach}[1]{\algorithmicforeach\ #1\ \algorithmicdo}

\doi{YQEA8075}

\makeatletter
\gdef\@copyrightpermission{
  \begin{minipage}{0.2\columnwidth}
   \href{https://creativecommons.org/licenses/by/4.0/}{\includegraphics[width=0.90\textwidth]{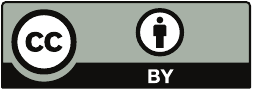}}
  \end{minipage}\hfill
  \begin{minipage}{0.8\columnwidth}
   \href{https://creativecommons.org/licenses/by/4.0/}{This work is licensed under a Creative Commons Attribution International 4.0 License.}
  \end{minipage}
  \vspace{5pt}
}
\makeatother

\setcopyright{ifaamas}
\acmConference[AAMAS '26]{Proc.\@ of the 25th International Conference
on Autonomous Agents and Multiagent Systems (AAMAS 2026)}{May 25 -- 29, 2026}
{Paphos, Cyprus}{C.~Amato, L.~Dennis, V.~Mascardi, J.~Thangarajah (eds.)}
\copyrightyear{2026}
\acmYear{2026}
\acmDOI{}
\acmPrice{}
\acmISBN{}

\title[AAMAS-2026 Formatting Instructions]{
Federated Gaussian Process Learning via\\Pseudo-Representations for Large-Scale Multi-Robot Systems 
}

\author{Sanket A. Salunkhe}
\affiliation{
  \institution{Colorado School of Mines}
  \city{Golden, CO}
  \country{USA}
  }
\email{sanket_salunkhe@mines.edu}

\author{George P. Kontoudis}
\affiliation{
  \institution{Colorado School of Mines}
  \city{Golden, CO}
  \country{USA}
  }
\email{george.kontoudis@mines.edu}

\begin{abstract}
Multi-robot systems require scalable and 
federated methods to model complex environments under computational and communication constraints. Gaussian Processes (GPs) offer robust probabilistic modeling, but suffer from cubic computational complexity, limiting their applicability in large-scale deployments. To address this challenge, we introduce the pxpGP, a novel distributed GP framework tailored for both centralized and decentralized large-scale multi-robot networks. Our approach leverages sparse variational inference to generate a local compact pseudo-representation. 
We introduce a sparse variational optimization scheme that bounds local pseudo-datasets and formulate a global scaled proximal-inexact consensus alternating direction method of multipliers (ADMM) with adaptive parameter updates and warm-start initialization. Experiments on synthetic and real-world datasets demonstrate that pxpGP and its decentralized variant, dec-pxpGP, outperform existing distributed GP methods in hyperparameter estimation and prediction accuracy, particularly in large-scale networks. 

\end{abstract}


\newcommand{\BibTeX}{\rm B\kern-.05em{\sc i\kern-.025em b}\kern-.08em\TeX}

\begin{document}


\pagestyle{fancy}
\fancyhead{}


\maketitle 

\noindent\textbf{Code:} \href{https://github.com/mpala-lab/distributed-gaussian-processes}{github.com/mpala-lab/distributed-gaussian-processes}


\section{Introduction}

Multi-robot systems are increasingly used
in executing complex, cooperative tasks such as environmental monitoring \cite{das2015data}, search-and-rescue \cite{queralta2020collaborative}, autonomous exploration~\cite{corah2019communication}, and surveillance~\cite{stump2011multi}. These applications require accurate modeling and prediction of environment or task-specific phenomena 
under uncertainty.~
Gaussian Processes (GPs)
are well suited to these challenges, combining accurate function approximation with explicit uncertainty quantification~\cite{rasmussen2006gaussian,gramacy2020surrogates}. 
They have been successfully applied to
distributed mapping~\cite{ghaffari2018gaussian,santos2021multi} and collaborative exploration tasks~\cite{suryan2020learning,kontoudis2021decentralized,kontoudis2025multi} due to their ability to provide uncertainty estimates that guide their decision-making process~\cite{luo2018adaptive,jang2020multi}. 

However, using GPs in multi-robot teams 
is constrained by practical challenges such as
limited onboard computation, privacy requirements, and communication bandwidth~\cite{gielis2022critical}. At the same time, GP training entails cubic complexity, 
which poses a major barrier with large datasets~\cite{kontoudis2023decentralized}. GP surrogate models are governed by a set of hyperparameters $\boldsymbol{\theta}$, learned using maximum likelihood estimation (MLE) methods over a given dataset~$\mathcal{D}$. Accurate hyperparameter estimation is essential to ensure 
reliable predictions~\cite{rasmussen2006gaussian}. Our objective in this work is to develop distributed GP learning methods that can accurately estimate GP hyperparameters in large-scale multi-robot systems without sharing local raw datasets.

GP approximation techniques can be broadly classified into 
global aggregation methods and local inducing point-based methods~\cite{liu2020gaussian}. 
Global approximation methods such as cGP \cite{xu2019wireless}, apxGP \cite{xie2019distributed}, and gapxGP \cite{kontoudis2024scalable} perform GP training across agents with Alternating Direction Method of Multipliers (ADMM) algorithms~\cite{boyd2011admm}. 
These approaches reduce computational and communication costs but require direct data sharing 
which can compromise data privacy and the quality of representation. Moreover, their performance degrades in networks larger than approximately $40$ agents due to the independent assumption of distributed optimization~\cite{kontoudis2024scalable,kontoudis2025multi}.

\begin{figure*}[!t]
\includegraphics[width=\textwidth]{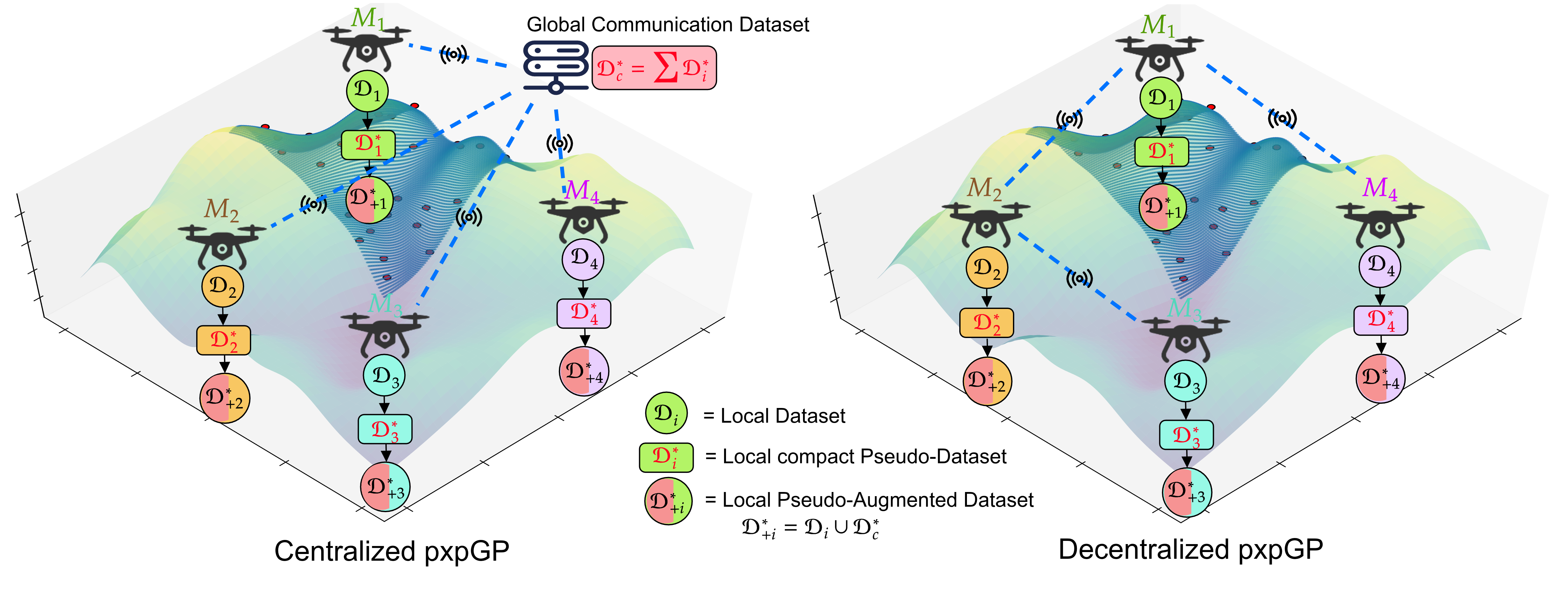}
 \captionof{figure}{Overview of the proposed pxpGP framework in centralized and decentralized multi-robot networks. Each agent~$M_i$ generates a compact pseudo-dataset 
 $\mathcal{D}_i^*$ and forms a pseudo-augmented dataset $\mathcal{D}_{+i}^*$. Centralized networks aggregate pseudo-datasets via a central node, while decentralized networks exchange data through neighbors via flooding. \label{fig:title}}
\end{figure*}

Local sparse variational methods 
reduce the cubic computational complexity of exact GPs by introducing a compact set of inducing variables that approximate the full covariance~\cite{titsias2009variational}. 
This method is infeasible in 
multi-robot systems, where data are inherently partitioned. To address this,~\cite{norton2022efficient} proposed a decentralized SGP framework, where each agent maintains a local variational posterior and fuses with neighboring models through maximum-consensus. 
While effective for small-scale deployments, this fusion mechanism is heuristic and lacks theoretical convergence guarantees, limiting scalability and robustness in large-scale networks. 

Recent efforts have explored adaptive sampling in multi-agent systems. In~\cite{brancato2024adaptive}, authors proposed a waypoint selection strategy where heterogeneous robots collaboratively estimate a stationary GP under dynamic constraints, and sensor noise. In~\cite{moreno2021modular}, the authors presented a centralized GP method using variational inference. Similarly, ~\cite{yue2024federated,llorente2025decentralized} develop decentralized GP approaches using random-feature GPs (RF-GPs), where agents share compact random-feature statistics with neighbors. However, random features can introduce systematic kernel approximation bias and yield poor covariance estimates, with accuracy strongly dependent on the number and quality of random features, in contrast to optimized pseudo-datasets.
Other works, such as COOL-GP~\cite{hoang2019collective} and mixture-of-experts-based adaptive sampling~\cite{masaba2023multi}, enable distributed GP learning and scalable modeling of non-stationary fields. While these methods advance distributed inference and adaptive data collection, they do not address the challenge of privacy-preserving hyperparameter optimization in GP training.



In this work, we propose the \textbf{\underline{P}roximal Ine\underline{x}act \underline{P}seudo Gaussian Process (pxpGP)}, a distributed GP training framework designed for large-scale centralized and decentralized multi-robot networks. The method lies at the intersection of global aggregation and local sparse approaches to achieve scalability and data privacy by exchanging only optimized pseudo-datasets among agents, rather than raw or random observations.
An overview of the proposed methods for both centralized and decentralized settings is illustrated in Fig.~\ref{fig:title}.

\subsubsection*{Contribution} The contribution of this work is twofold. First, we extend sparse variational inference techniques \cite{snelson2005sparse,norton2023decentralized,green2024distributed} to generate \textit{compact pseudo-datasets} 
confined to each agent’s region, improving informativeness and scalability to large-scale networks. 
Federated learning is promoted by sharing only compact pseudo-representations and optimization iterates instead of raw data.
Second, we formulated pxpGP as a \textit{scaled proximal-inexact consensus ADMM} algorithm 
initialized with warm-start hyperparameters and adaptive residual balancing
that accelerates convergence and reduces communication rounds.
\section{Gaussian Process Training}\label{sec:probForm}

Gaussian Processes (GPs) are non-parametric Bayesian models that 
define distributions over functions with a Gaussian prior. A GP over a latent function $f(\boldsymbol{x})$ is defined as,
\begin{equation*}
\begin{aligned}
    f(\boldsymbol{x}) \sim  GP(m(\boldsymbol{x}), k(\boldsymbol{x}, \boldsymbol{x}^{\prime})),   
\end{aligned}
\label{eq:gp_equation}
\end{equation*}
where $m(\boldsymbol{x})$ is the mean function and $k(\boldsymbol{x}, \boldsymbol{x}^{\prime})$ is the covariance function (i.e., kernel), parameterized by a set of hyperparameters $\boldsymbol{\theta}$, that govern the smoothness, variability, and predictive accuracy of the GP model.



We model observations as $y(\boldsymbol{x}) = f(\boldsymbol{x}) + \epsilon$, where $\boldsymbol{x} \in \mathbb{R}^D$ is the input with dimension $D$, $y(\boldsymbol{x}) \in \mathbb{R}$ is the scalar output, and $\epsilon \sim \mathcal{N}(0, \sigma_{\epsilon}^2)$ is zero-mean Gaussian measurement noise with variance $\sigma_{\epsilon}^2$.
We employ the {Separable Squared Exponential (SSE)} kernel,
\begin{equation*}
\begin{aligned}
    k(\boldsymbol{x} ,\boldsymbol{x}^{\prime}) = \sigma_f^2 \exp \left\{ -   \frac{1}{2} \sum_{d=1}^D\frac{\left({x}_d - {x}_d^{\prime}\right)^2 }{l_d^2}\right\},\end{aligned}
\label{eq:sse_kernel}
\end{equation*}
with signal variance $\sigma_f > 0$ and length-scale $l_d > 0$. The 
GP hyperparameters $\boldsymbol{\theta} = 
\left[ l_1, l_2, \cdots, l_d, \sigma_{f}, \sigma_{\epsilon} \right]^T \in \mathbb{R}^{D+2}$ are trained by maximizing the log-
likelihood function, 
\begin{equation*}
\begin{aligned}
    \mathcal{L}(\boldsymbol{X},\boldsymbol{y}; \boldsymbol{\theta}) = 
    -\frac{1}{2} \left( \boldsymbol{y}^{\intercal}\boldsymbol{C}_{\theta}^{-1}\boldsymbol{y} 
        + \log |\boldsymbol{C}_{\theta}| +N\log 2\pi \right),
\end{aligned}
\label{eq:mll_equation}
\end{equation*}
where $\boldsymbol{C}_{\theta} = \boldsymbol{K}+\sigma_{\epsilon}^2I_N$ is the positive definite covariance matrix and $\boldsymbol{K} = k(\boldsymbol{X},\boldsymbol{X})$ is the kernel matrix, 
with $\boldsymbol{X} = \left\{ \boldsymbol{x}_1, \boldsymbol{x}_2, \cdots, \boldsymbol{x}_n \right\}_{n=1}^N \subset \mathbb{R}^{N\times D}$ the input locations, $\boldsymbol{y} = \left\{ {y}_1, {y}_2, \cdots, {y}_n \right\}_{n=1}^N \subset \mathbb{R}^{N}$ the corresponding scalar outputs, and $N$ the dataset size. 
Thus, the negative log-likelihood (NLL) optimization problem yields,
\begin{align} \label{eq:nll_min_problem}
    \hat{\boldsymbol{\theta}} = \arg\min_{\boldsymbol{\theta}} \quad & \boldsymbol{y}^{\intercal} \boldsymbol{C}_{\theta}^{-1} \boldsymbol{y} 
        +  \log |\boldsymbol{C}_{\theta}| \\ \nonumber
    \text{s.t.} \quad & \boldsymbol{\theta} > \boldsymbol{0}_{D+2}.
\end{align}
The positivity constraint keeps 
$\boldsymbol{C}_{\theta}$ well-conditioned and positive definite. The optimization 
\eqref{eq:nll_min_problem} requires computing $\boldsymbol{C}_{\theta}^{-1}$ at each iteration, with $\mathcal{O}(N^3)$ computations and $\mathcal{O}(N^2 + DN)$ storage.

\begin{figure*}[!t]
    \centering
    \begin{subfigure}[t]{0.32\textwidth}
        \includegraphics[width=\linewidth]{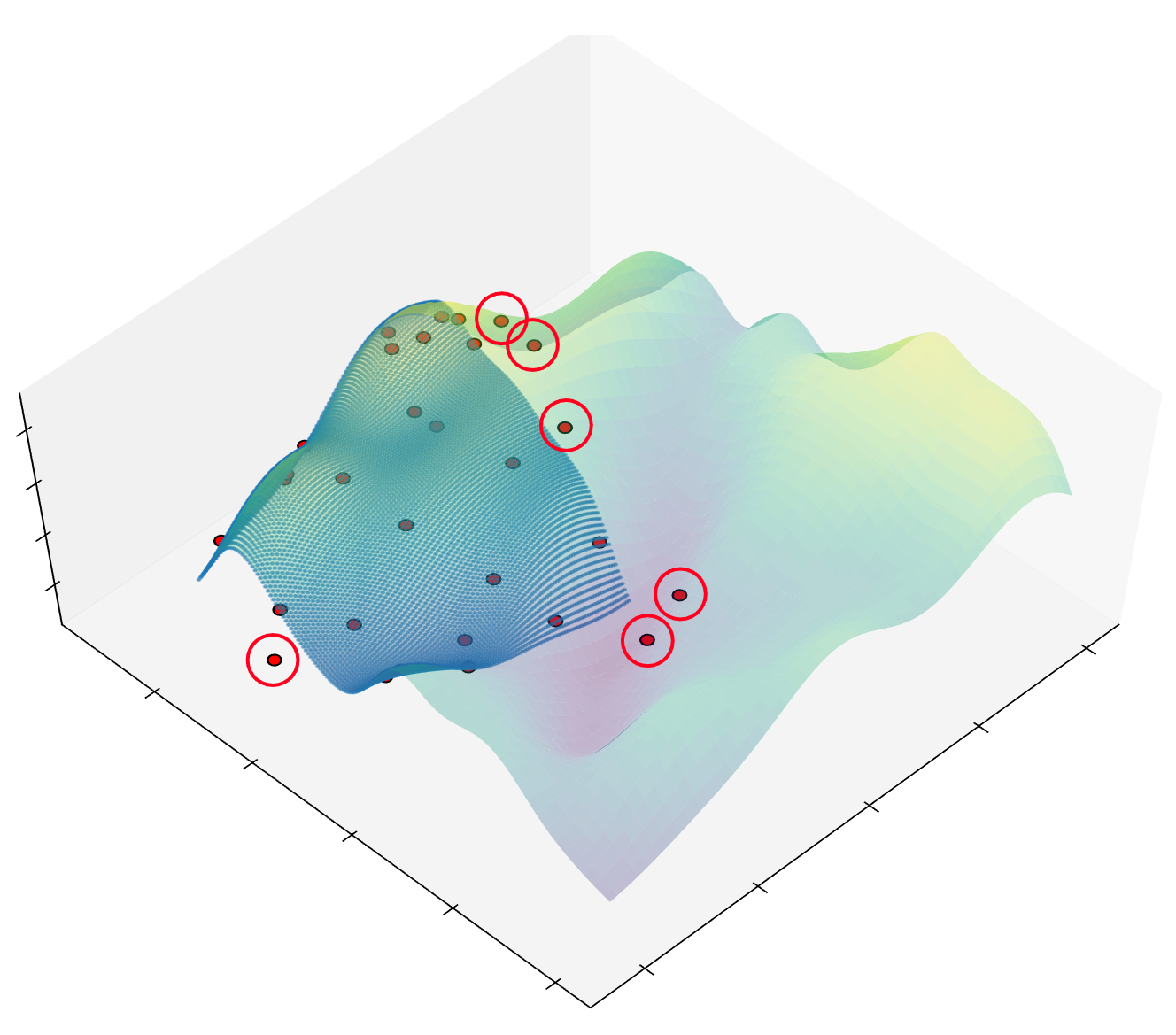}
        \caption{Without Boundary Penalty}
        \label{fig:penalty_b}
    \end{subfigure}
    \hfill
    \begin{subfigure}[t]{0.32\textwidth}
        \includegraphics[width=\linewidth]{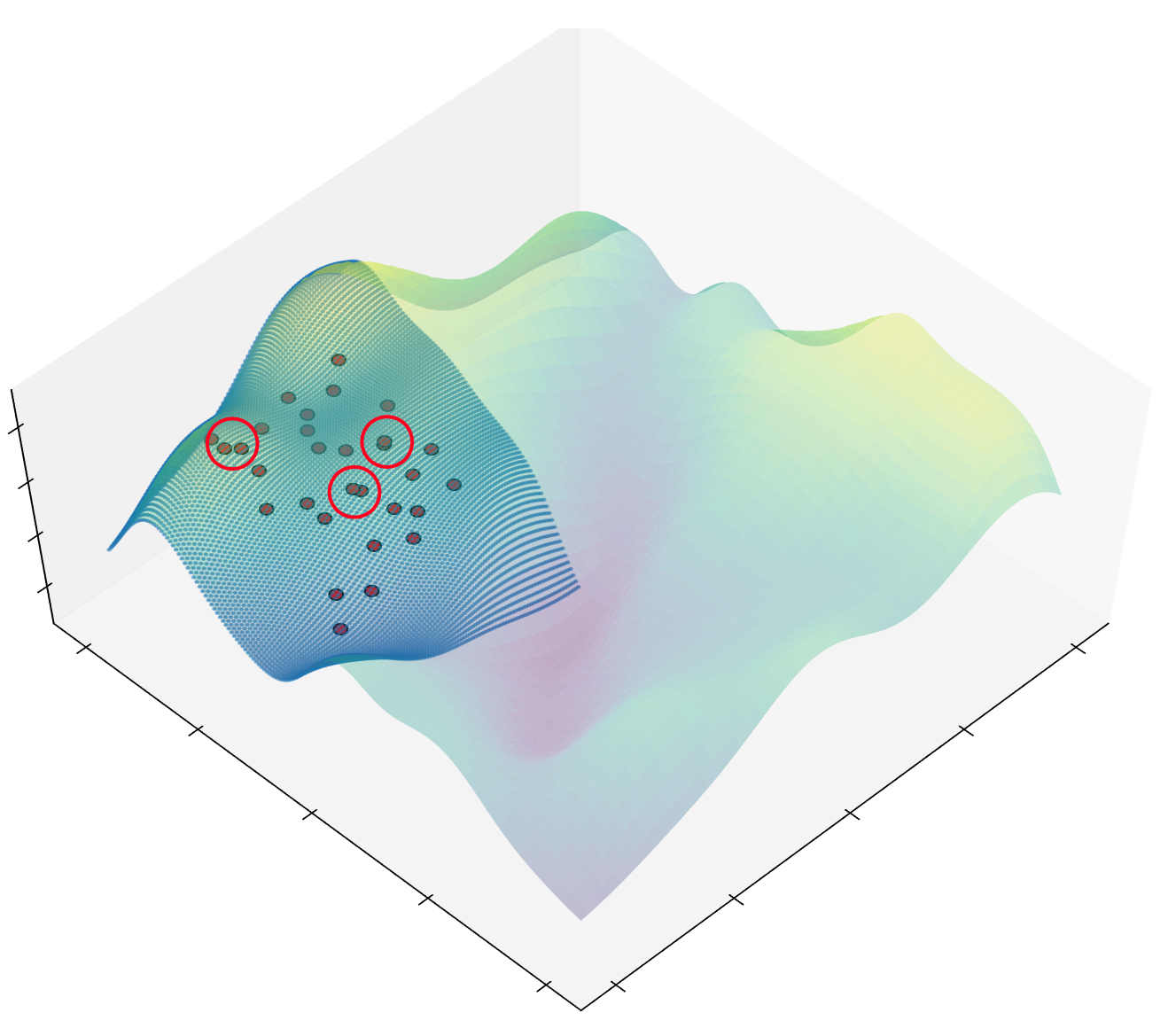}
        \caption{Without Repulsive penalty}
        \label{fig:penalty_r}
    \end{subfigure}
    \hfill
    \begin{subfigure}[t]{0.32\textwidth}
        \includegraphics[width=\linewidth]{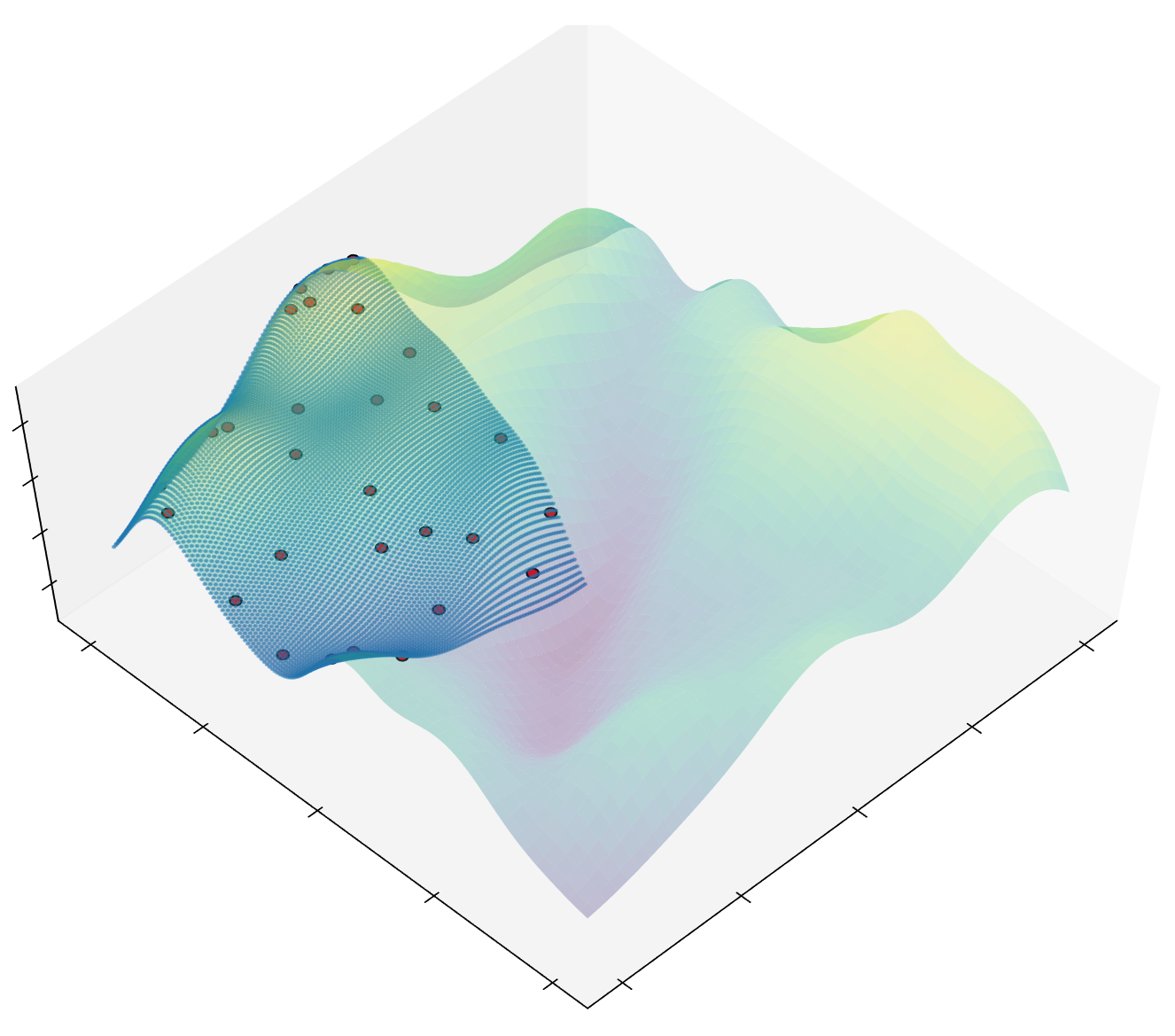}
        \caption{With Boundary \& Repulsive Penalty}
        \label{fig:penalty_w}
    \end{subfigure}
    \caption{Effect of pxpGP regularization. (a) Pseudo-points drift beyond local bounds without boundary penalty (highlighted red circles). (b) Without the repulsive penalty, points cluster densely in a local region (highlighted red circles). (c) Combined boundary ($\mathfrak{L}_{b}$) and repulsive ($\mathfrak{L}_{r}$) penalties yield a well-distributed local pseudo-representations. }
    \label{fig:penalty_combined} 
\end{figure*}

\subsection{Centralized Factorized GP Training (fact-GP)}
To reduce the complexity of GP training, factorized GP (fact-GP) \cite{deisenroth2015distributed} partitions the global dataset $\mathcal{D}=\left\{ \boldsymbol{X, y} \right\}$ across multiple agents $M$ disjoint subsets, $\mathcal{D}= \{ \mathcal{D}_i\}_{i=1}^M$, where $i = \left\{ 1, 2, \cdots, M \right\}$ and $\mathcal{D}_i = \left\{ \boldsymbol{X}_i, \boldsymbol{y}_i \right\}$.~
The global objective is approximated by the sum of local objectives with independent local datasets,~
$\mathcal{L} \approx \sum_{i=1}^{M} \mathcal{L}_i$.~
Each agent $i$ trains local hyperparameters~$\boldsymbol{\theta}_i$
and enforces global consensus through a shared parameter~$\boldsymbol{z}$, 
\begin{align}\label{eq:admm_nll_min_problem}
    \hat{\boldsymbol{\theta}} =& \arg\min_{\boldsymbol{\theta}} \sum_{i=1}^{M} \mathcal{L}_i(\boldsymbol{\theta}_i)   \\ \nonumber
    \text{s.t.} \quad & \boldsymbol{\theta}_i = \boldsymbol{z}, \quad \forall i = 1, 2, \cdots, M,
\end{align}
where $\mathcal{L}_i(\boldsymbol{\theta}_i) = \boldsymbol{y}_i^{\intercal} \boldsymbol{C}_{\theta ,i}^{-1} \boldsymbol{y}_i +  \log |\boldsymbol{C}_{\theta,i}|$ is the local NLL and $\boldsymbol{C}_{\theta,i}$ is the local covariance matrix. 
To formulate the proposed distributed training algorithms, we introduce two specific assumptions 
about data distribution and communication structure among agents.

\begin{assumption}\label{assumption:dist_dataset}
Each agent $i$ trains a local sub-model 
on a statistically independent dataset that corresponds to a distinct region of the input space.
\end{assumption}

\begin{assumption}\label{assumption:federated_learning}
Communication between agents is restricted to parameter or summary 
exchange and does not involve sharing raw datasets to preserve data privacy.
\end{assumption}

The approximate proximal GP (apx-GP)~\cite{xie2019distributed} uses proximal inexact consensus ADMM to solve~\eqref{eq:admm_nll_min_problem}, reducing local complexity to $\mathcal{O}((N/ M)^3)$ with convergence guarantees for the non-convex optimization. It enables GP models to scale over large datasets,~
but as the number of agents $M$ increases, Assumption~\ref{assumption:dist_dataset} weakens, degrading 
hyperparameter estimates. On the other hand, gapx-GP~\cite{kontoudis2024scalable} addresses this challenge by augmenting each local dataset $\mathcal{D}_{+i}$ with randomly sampled data from other agents. 
However, the latter violates Assumption~\ref{assumption:federated_learning} about privacy and does not scale beyond networks of approximately 40 agents.

\subsection{Decentralized GP Training}
In practical scenarios, a central coordinator is infeasible due to communication constraints, which motivates decentralized GP training where each agent collaborates only with its immediate neighbors~$\mathcal{N}_i$.
We model the decentralized network of $M$ agents as a connected undirected graph $\mathcal{G} = \left( \mathcal{V}, \mathcal{E} \right)$, with $\mathcal{V} = \left\{ v_1, v_2, \cdots, v_M \right\}$ as a set of agents i.e nodes in the network, and $\mathcal{E} \subseteq \mathcal{V} \times \mathcal{V}$ the communication links i.e edges between them. For each agent $i$, the set of neighbors 
is defined as $\mathcal{N}_i = \left\{ v_j \in \mathcal{V} | \left( v_i, v_j \right) \in \mathcal{E} \right\}$. 

Prior work~\cite{kontoudis2024scalable} introduced dec-cGP, dec-apxGP, and dec-gapxGP, 
to decentralized networks using edge-based ADMM~\cite{shi2014linear} to solve,
\begin{align}\label{eq:dec_admm_nll_min_problem}
    \hat{\boldsymbol{\theta}} =& \arg\min_{\boldsymbol{\theta}} \sum_{i=1}^{M} \mathcal{L}_i(\boldsymbol{\theta}_i)   \\ \nonumber
    \text{s.t.} \quad & \boldsymbol{\theta}_i = \boldsymbol{z}_{ij}, \quad \forall i \in \mathcal{V}, j \in \mathcal{N}_i \\ \nonumber
    & \boldsymbol{\theta}_j = \boldsymbol{z}_{ij}, \quad \forall i \in \mathcal{V}, j \in \mathcal{N}_i,
\end{align}
where each agent optimizes its local hyperparameters $\boldsymbol{\theta}_i$ while maintaining consensus with neighbors via shared auxiliary variables $\boldsymbol{z}_{ij}$. The per-agent complexity remains 
$\mathcal{O}\left((N/M)^3\right)$. Similar to the gapxGP, dec-gapxGP also shares raw data with neighboring agents, violating Assumption \ref{assumption:federated_learning}. 

\begin{problem}
    Consider a large network of $M$ robots that 
    collaboratively model an unknown latent function using GPs. Each agent $i$ holds 
    a local dataset $\mathcal{D}_i$ and communicates with its one-hop neighbors. Under Assumption~\ref{assumption:dist_dataset} (Independence) and~\ref{assumption:federated_learning} (Federated Constraints), the goal is to estimate the global GP hyperparameters $\boldsymbol{\hat{\theta}}$ by solving the centralized optimization problem~\eqref{eq:admm_nll_min_problem} and its decentralized counterpart~\eqref{eq:dec_admm_nll_min_problem}, 
    while minimizing communication rounds by ensuring fast convergence. 
\end{problem}

\section{Pseudo Inexact Proximal GP (pxp-GP) Training}\label{sec:decentTrain}

In this section, we present the formulation of the \textbf{Proximal Inexact Pseudo GP (pxpGP)} training method for centralized and decentralized networks. 
Existing distributed GP training methods, such as gapxGP and dec-gapxGP \cite{kontoudis2024scalable}, reduce approximation error of factorized GP methods by augmenting each agent’s dataset~$\mathcal{D}_{+i}$ with randomly sampled data from other agents. While these approaches are effective for small networks, they i) yield poorly representative augmented datasets in large-scale networks and ii) require raw-data exchange that violates the federated constraint (Assumption~\ref{assumption:federated_learning}).
pxpGP addresses these issues by letting each agent build a local pseudo-augmented dataset $\mathcal{D}_{+i}^*$ from a local sparse GP model trained over its local dataset $\mathcal{D}_i$, rather than from random samples. 


Sparse GP approximations use a compact set of $P$ inducing points $\boldsymbol{X}_p = \left\{ x_{p_1}, x_{p_2}, \cdots, x_{p_i} \right\}_{i=1}^P \subset \mathbb{R}^{P\times D}$, where typically $P << (N/M)$. Thus, the training and prediction complexity reduces to $\mathcal{O}(NP^2)$ and $\mathcal{O}(P^2)$, respectively~\cite{titsias2009variational, snelson2005sparse}. The inducing points $\boldsymbol{X}_p$, variational parameters $\boldsymbol{\mu}_P$, and $\boldsymbol{A}_p$ are optimized by minimizing the negative Evidence Lower Bound (ELBO) $\mathfrak{L}_{\text{ELBO}}$ that yields,
\begin{align}\nonumber
    &q(f_P) = \min_{q(f_P), \boldsymbol{X}_P} - \mathfrak{L}_{\text{ELBO}}\\ \label{eq:min_elbo_divergence}
    &= \min_{q(f_P), \boldsymbol{X}_P} - \left( \mathbb{E}_{q(\boldsymbol{f})} \left[ \text{log } p(\boldsymbol{y|f})\right]  - \text{KL} \left( q(\boldsymbol{f}_P) \mid \mid p(\boldsymbol{f}_P) \right) \right), 
\end{align}
where $q(f_P) = \mathcal{N} \left( \boldsymbol{\mu}_P, \boldsymbol{A}_P \right)$ is a Gaussian distribution with $\boldsymbol{\mu}_P$ variational mean and $\boldsymbol{A}_P$ variational covariance matrix, 
and $\text{KL} \left( q \mid \mid p \right)$ is the Kullback-Leibler (KL) divergence between the variational distribution $q$ and GP prior $p$ over the inducing variables $\boldsymbol{f}_P$.
The first term of $\mathfrak{L}_{\text{ELBO}}$ encourages accurate data fitting, and the second term regularizes the variational approximation by penalizing divergence from the true posterior. The optimization starts with K-means initialization of the inducing points $\boldsymbol{X}_P$, followed by variational inference to minimize the negative ELBO~\eqref{eq:min_elbo_divergence}.

Once each agent $i$ generates a local pseudo-dataset $\mathcal{D}_{i}^*$ by solving~\eqref{eq:min_elbo_divergence}, then all agents transmit $\mathcal{D}_{i}^*$ to 
create a shared communication dataset $\mathcal{D}_{c}^* = \cup_{i=1}^M \mathcal{D}_i^*$. Next, each agent constructs its local pseudo-augmented dataset $\mathcal{D}_{+i}^* = \mathcal{D}_{i} \cup  \mathcal{D}_{c}^*$ by merging the original local dataset with the shared communication dataset, providing a richer global representation for GP training.
The quality of the pseudo-dataset largely depends on the placement of inducing points. Without constraints, inducing points may drift beyond local data boundaries or cluster in dense regions as shown in Fig.~\ref{fig:penalty_b}, \ref{fig:penalty_r}, respectively, leading to poor generalization, ill-conditioned covariance matrices, and occasionally Cholesky decomposition failures.
To mitigate this, we introduce two regularization terms in the variational ELBO: 1) a boundary penalty ($\mathfrak{L}_{b}$) to confine points within data bounds; and 2)~a repulsive penalty ($\mathfrak{L}{r}$) to ensure well-spread inducing points.



\subsubsection{Boundary Penalty ($\mathfrak{L}_{b}$)} 
This penalty constrains inducing points to remain within the bounds of the local dataset,
\begin{align}\label{eq:boundary_penalty}
    \mathfrak{L}_{b} = \sum_{i = 1}^{P} \text{ReLU} \left( \boldsymbol{x}_{\text{min}} - \boldsymbol{x}_{i}^* \right)^2 + \text{ReLU} \left( \boldsymbol{x}_{i}^* - \boldsymbol{x}_{\text{max}} \right)^2 ,
\end{align}
where $\boldsymbol{x}_{i}^*$ represents the $i$-th inducing pseudo-input point among $P$ points, $\boldsymbol{x}_{\text{min}}$ and $\boldsymbol{x}_{\text{max}}$ denote the minimum and maximum boundaries of the local dataset, respectively. The Rectified Linear Unit (ReLU) function ensures zero penalty inside the valid region, but applies a quadratic cost when points stray beyond the 
boundaries.




\subsubsection{Repulsive Penalty ($\mathfrak{L}_{r}$)} 
The variational sparse GP objective~\eqref{eq:min_elbo_divergence} \cite{titsias2009variational} inherently discourages redundant overlapping inducing points through its complexity and trace terms. However, this implicit repulsive and non-overlapping effect is soft and data-dependent. Thus, the variational sparse GP objective~\eqref{eq:min_elbo_divergence} does not explicitly prevent local clustering, especially in distributed, data-partitioned, or non-stationary multi-agent setups where data exhibit spatial bias. As a result, the standard ELBO may yield clustered or poorly spaced inducing points, as seen in Fig.~\ref{fig:penalty_r}. To address this limitation, the proposed repulsive penalty $\mathfrak{L}_r$ introduces an explicit geometric prior to enforce a minimum separation distance $d_{\text{min}}$ between pseudo-inputs and improve spatial coverage, 
\begin{align}\label{eq:replusive_penalty}
    \mathfrak{L}_{r} = \sum_{i = 1}^{P} \sum_{j = 1}^{P} \text{ReLU} \left( d_{\text{min}} - \left\|  \boldsymbol{x}_{i}^* - \boldsymbol{x}_{j}^* \right\| \right)^2 ,
\end{align}
where $\left\| \cdot \right\|$ denotes the Euclidean norm between two inducing points. The ReLU function ensures zero penalty when points are sufficiently separated, but applies a quadratic cost distance between points that fall below the threshold $d_{\text{min}}$.

Together, these penalties produce a compact, well-distributed, and privacy-preserving local pseudo-augmented dataset $\mathcal{D}_{+i}^*$ that enhances global GP approximation and improves numerical conditioning of covariance matrices.
The final objective function for Sparse GP combines ELBO~\eqref{eq:min_elbo_divergence} with the boundary~\eqref{eq:boundary_penalty} and repulsive penalties~\eqref{eq:replusive_penalty} as,
\begin{align}\label{eq:elbo_final_obj}
    \boldsymbol{X}_P =&\argmin_{q(f_P), \boldsymbol{X}_P} 
    - \mathfrak{L}_{\text{ELBO}} + \mathfrak{L}_{b} + \mathfrak{L}_{r}.
\end{align}



\subsection{Centralized pxpGP training} 
In the proposed centralized pxpGP framework, each agent~$i$ optimizes the hyperparameters $\boldsymbol{\theta}_i$ of its local GP model using the local pseudo-augmented dataset $\mathcal{D}_{+i}^*$. This is formulated as a scaled proximal-inexact consensus ADMM (pxADMM) problem, with analytical 
synchronous iterates~\cite{hong2016convergence} coordinated by a central node, with a fixed set of participating agents. By introducing a scaled dual variable $u_i^k = \frac{1}{\rho}\lambda_i^k$, we simplify the update rules, improve numerical stability, and enable adaptive penalty updates and warm-start initialization. 

\newfloat{algorithm}{!t}{lop}
\begin{algorithm}
\caption{pxpGP}
\begin{algorithmic}[1]

\Statex \textbf{Input:} $\mathcal{D}_i = (\boldsymbol{X}_i, \boldsymbol{y}_i)$, $k(\cdot,\cdot)$, $\rho_i$, $L_i$, $\epsilon_{\text{abs}}$, $\epsilon_{\text{rel}}$ 

\Statex \textbf{Output:} $\hat{\boldsymbol{\theta}}$, $\mathcal{D}_{+i}^*$

\For{$i = 1$ to $M$}  \Comment{Sparse Modeling}
    \State $\mathcal{D}_i^*, \boldsymbol{\theta}_{i}^* \leftarrow \texttt{SparseModel}(\mathcal{D}_i) $ \eqref{eq:elbo_final_obj}
    \State Communicate $\mathcal{D}_i^*$ to central node.
\EndFor

\State Aggregate $\mathcal{D}_c^* = \cup_{i=1}^M \mathcal{D}_i^*$ at central node.
\State Broadcast $\mathcal{D}_c^*$ to all agents $i\in M$ from central node.

\For{$i = 1$ to $M$}  
    \State $\mathcal{D}_{+i}^* = \mathcal{D}_{i} \cup  \mathcal{D}_{c}^*$  \Comment{Local Augmented Dataset}
    \State Initialize $\boldsymbol{\theta}_i^{(1)} = \boldsymbol{\theta}_{i}^*$ \Comment{Warm Start}
\EndFor

\Repeat \Comment{ADMM optimization}
    \State Communicate $\boldsymbol{\theta}_i^{(s)}$ to central node.
    \State $\boldsymbol{z}^{(s+1)} \leftarrow \texttt{primal-2}(\boldsymbol{\theta}_i^{(s)}, \boldsymbol{u}_i^{(s)})$ (\ref{eq:pxpgp_z_update}) 
    \State Broadcast $\boldsymbol{z}^{(s+1)}$ to all agents from central node.

    \For{$i = 1$ to $M$}  
        \State $\boldsymbol{\theta}_i^{(s+1)} \leftarrow  \texttt{primal-1}(\boldsymbol{z}^{(s+1)}, \boldsymbol{u}_i^{(s)}, \mathcal{D}_{+i}^*)$ (\ref{eq:pxpgp_theta_update})
        \State $\boldsymbol{u}_i^{(s+1)} \leftarrow \texttt{dual}(\boldsymbol{u}_i^{(s)}, \boldsymbol{\theta}_i^{(s+1)}, \boldsymbol{z}^{(s+1)})$ (\ref{eq:pxpgp_u_update})
        \State Update $\rho_i^{(s+1)}$ (\ref{eq:adaptive_rho}), $L_i^{(s+1)}$ (\ref{eq:adaptive_lip})
    \EndFor
    
\Until{$\left\| \boldsymbol{r}_i^{(s+1)} \right\| \le \epsilon_{\text{primal}}$~\eqref{eq:convergence_primal}, $\left\| \boldsymbol{s}_i^{(s+1)} \right\| \le \epsilon_{\text{dual}}$~\eqref{eq:convergence_dual} }

\State \Return $\hat{\boldsymbol{\theta}}$
\end{algorithmic}
\label{algo:cent_pxpgp}
\end{algorithm}

Consistent with existing distributed GP training methods such as cGP \cite{xu2019wireless}, apxGP \cite{xie2019distributed}, and gapxGP \cite{kontoudis2024scalable}, pxpGP also enables each local agent to train independently while maintaining global consensus 
\eqref{eq:admm_nll_min_problem}.
The pxADMM linearizes the augmented Lagrangian around a stationary point $\boldsymbol{v}_i = \boldsymbol{z} + \boldsymbol{u}_i$ that yields, 
\begin{align}\nonumber
        \mathscr{L} \left(  \boldsymbol{ \theta}_i , \boldsymbol{z}, \boldsymbol{v}_i \right) &=  \sum_{i=1}^{M} \mathcal{L}_i(\boldsymbol{z})+ \nabla_{\boldsymbol{\theta}}^{\intercal} \mathcal{L}_i(\boldsymbol{v}_i) \left( \boldsymbol{\theta}_i - \boldsymbol{v}_i \right) \\ \label{eq:augmented_lagrangian_equation}
        & \quad + 
        \frac{L_i + \rho_i}{2} \left\|  \boldsymbol{\theta}_i  - \boldsymbol{v}_i \right\|^2,
\end{align}
where $L_i > 0$ is a positive Lipschitz parameter and $\rho_i$ a regularization penalty parameter.
The iterative updates for the pxpGP are provided by,
\begin{subequations}
\begin{align}
    \boldsymbol{\theta}_i^{(s+1)} &= \boldsymbol{v}_i^{(s)} - \frac{1}{L_i^{(s)} + \rho_i^{(s)}} \nabla_{\boldsymbol{\theta}} \mathcal{L}_i\left(\boldsymbol{v}_i^{(s)}\right) \label{eq:pxpgp_theta_update} \\
    \boldsymbol{z}^{(s+1)} &= \frac{1}{M} \sum_{i=1}^{M} \left( \boldsymbol{\theta}_i^{(s+1)} + \boldsymbol{u}_i^{(s)} \right) \label{eq:pxpgp_z_update} \\
    \boldsymbol{u}_i^{(s+1)} &= \boldsymbol{u}_i^{(s)} + \boldsymbol{\theta}_i^{(s+1)} - \boldsymbol{z}^{(s+1)}. \label{eq:pxpgp_u_update}
\end{align}
\label{eq:pxpgp_iterative_scheme}
\end{subequations}

While optimizing the global hyperparameters $\boldsymbol{\theta}$, the proposed pxpGP framework leverages the locally learned variational hyperparameters $\boldsymbol{\theta}_i^*$ from each sparse GP model to initialize subsequent global training rounds. This warm-start mechanism preserves posterior information from local models and accelerates convergence by providing informed initial estimates for $\boldsymbol{\theta}_i^{(1)}=\boldsymbol{\theta}_i^*$.

\begin{figure*}[!t]
    \centering
    \begin{subfigure}[t]{0.20\textwidth}
        \centering
        \includegraphics[width=\linewidth]{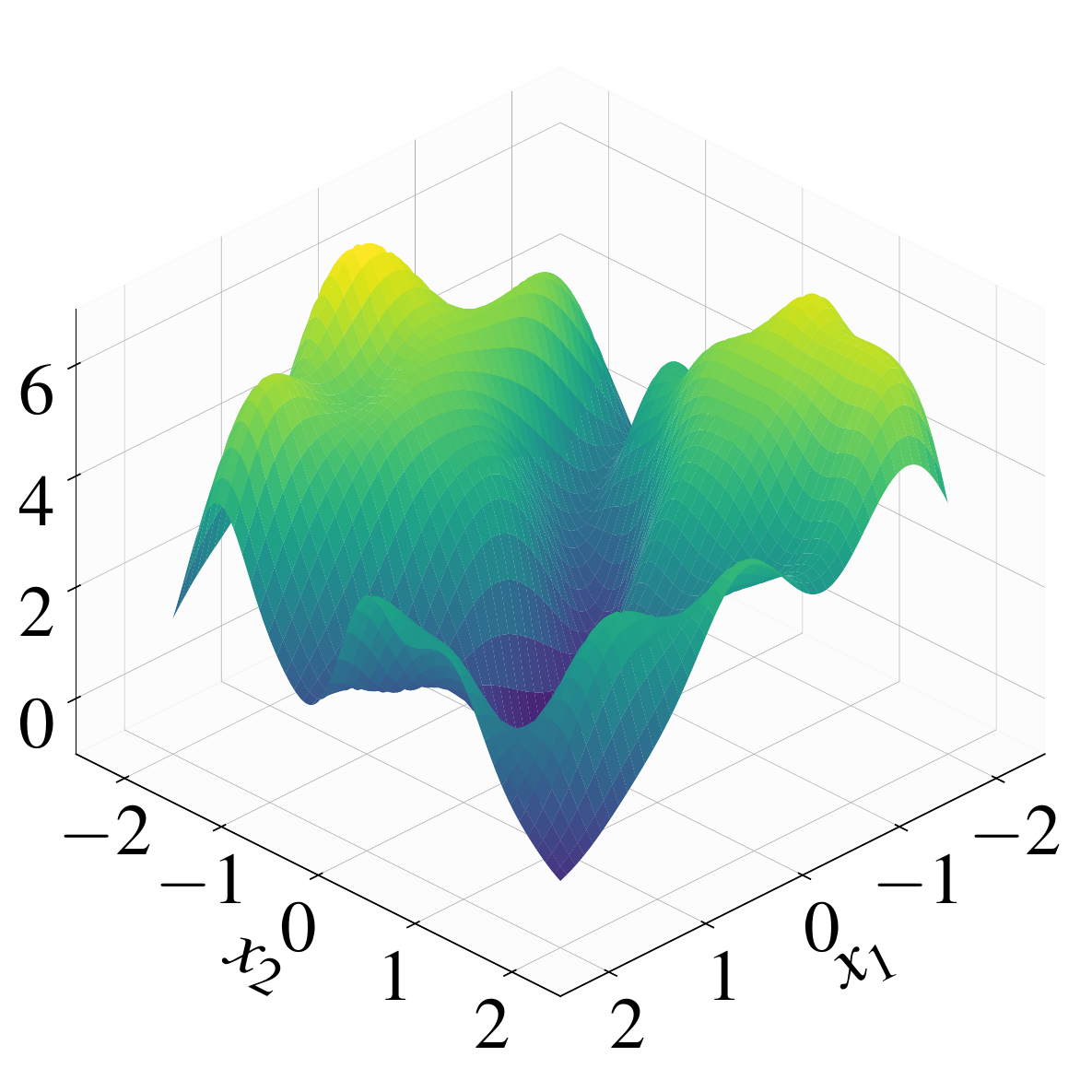}
        \caption{GP Generative 1}
        \label{fig:dataset_1}
    \end{subfigure}
    \hfill
    \begin{subfigure}[t]{0.20\textwidth}
        \centering
        \includegraphics[width=\linewidth]{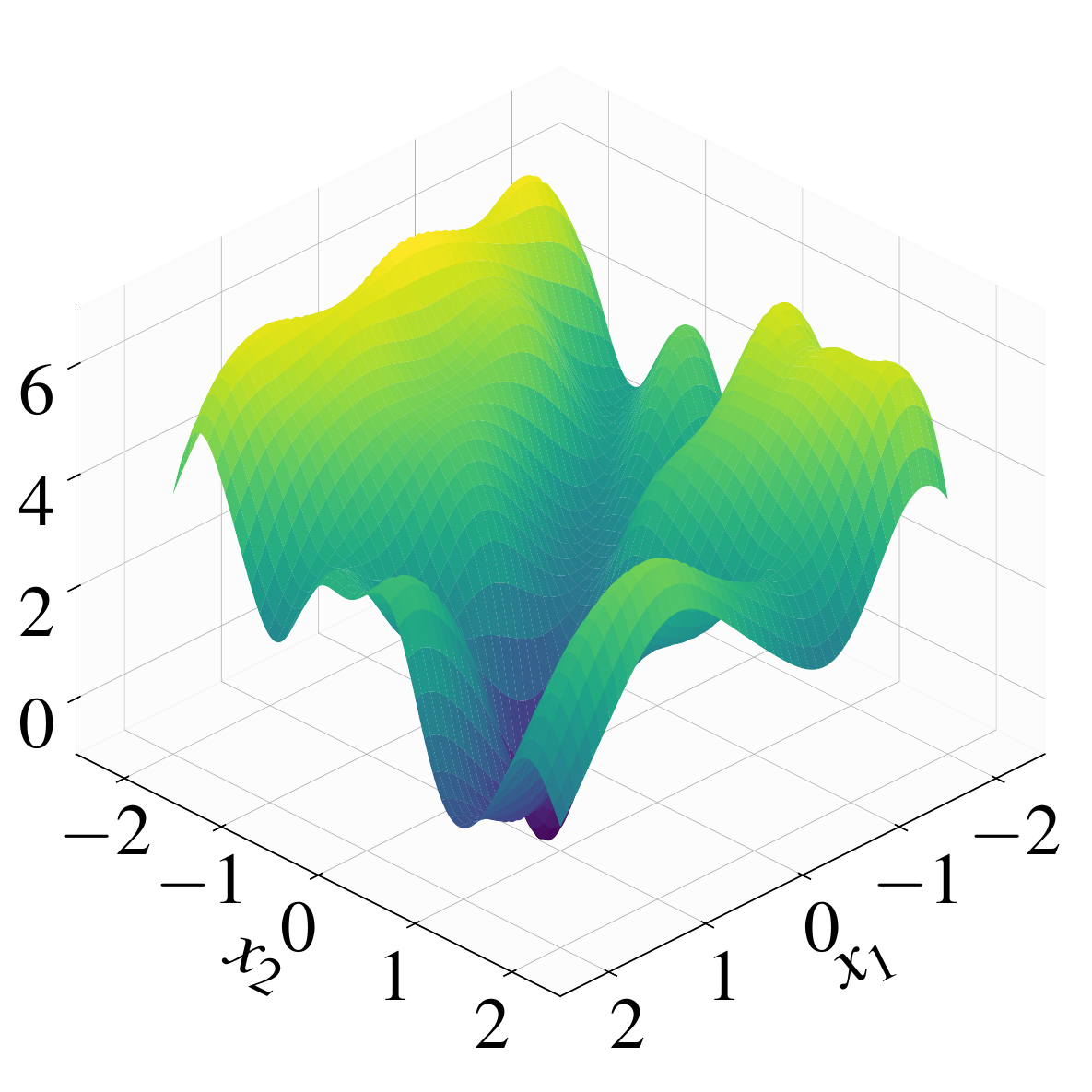}
        \caption{GP Generative 2}
        \label{fig:dataset_2}
    \end{subfigure}
    \hfill
    \begin{subfigure}[t]{0.20\textwidth}
        \centering
        \includegraphics[width=\linewidth]{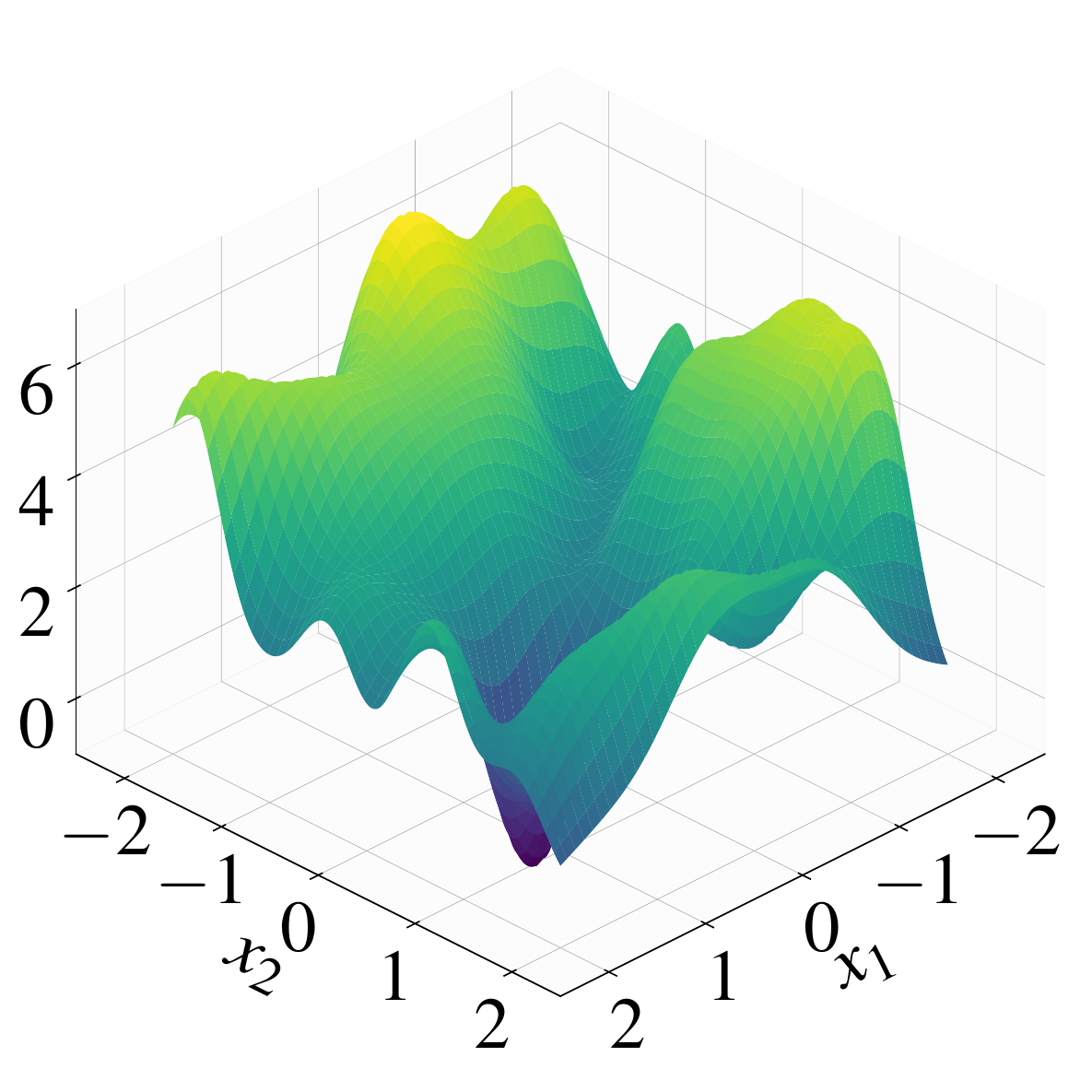}
        \caption{GP Generative 3}
        \label{fig:dataset_3}
    \end{subfigure}
    \hfill
    \begin{subfigure}[t]{0.18\textwidth}
        \centering
        \includegraphics[width=\linewidth]{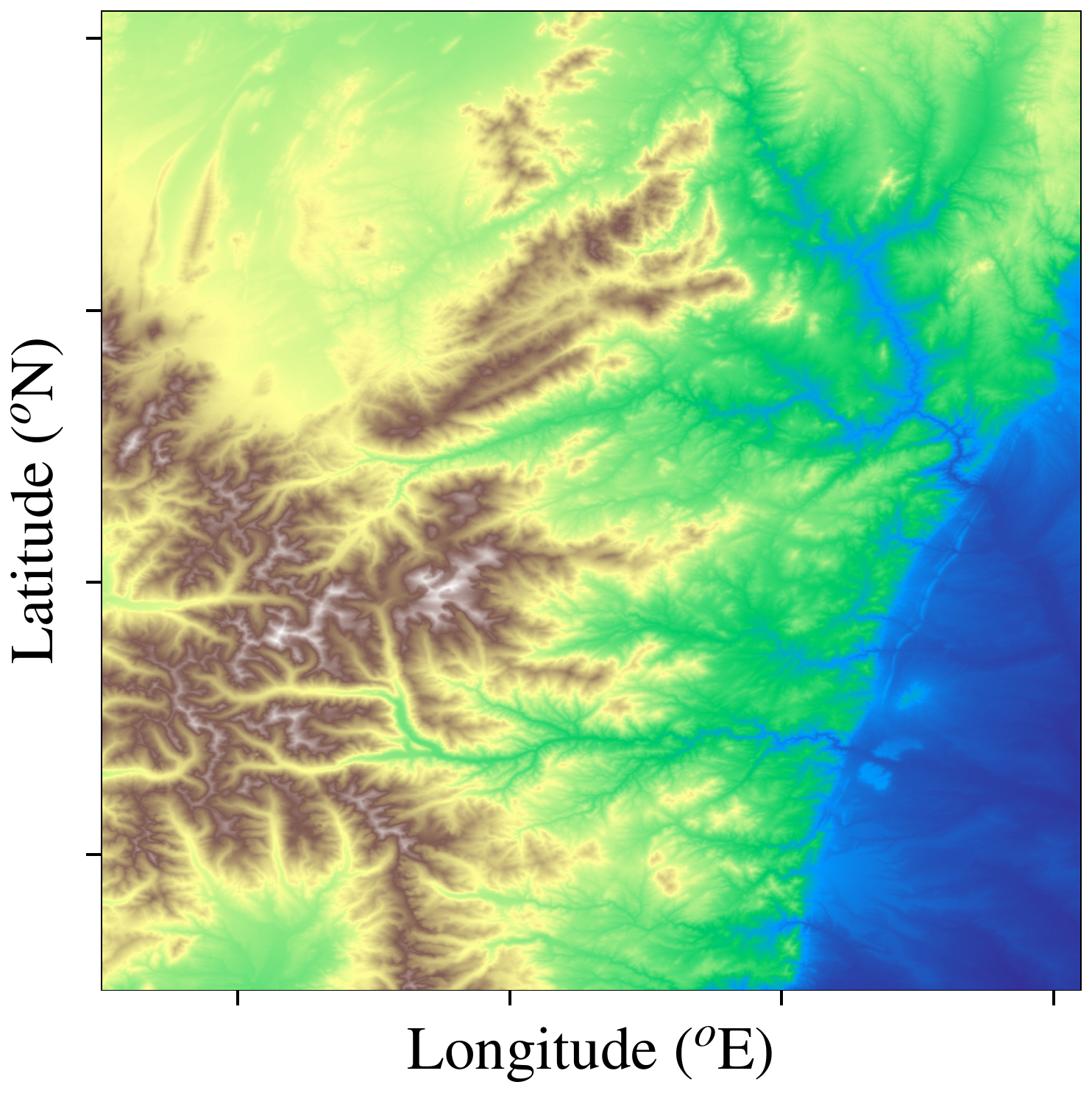}
        \caption{N39W106}
        \label{fig:dataset_4}
    \end{subfigure}
    \hfill
    \begin{subfigure}[t]{0.18\textwidth}
        \centering
        \includegraphics[width=\linewidth]{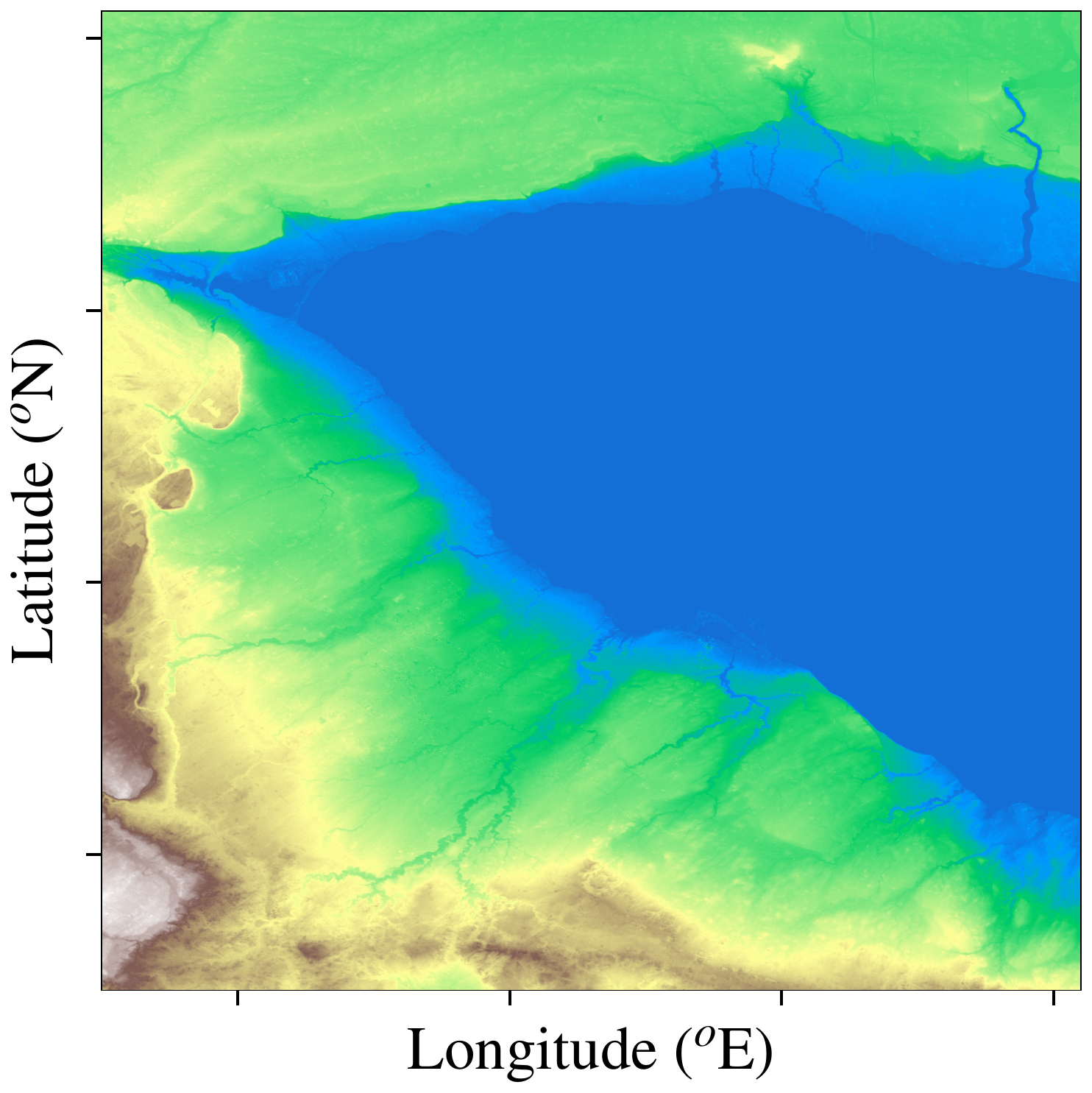}
        \caption{N43W080}
        \label{fig:dataset_5}
    \end{subfigure}

    \caption{Visualization of the datasets used for experimentation. Figures~\eqref{fig:dataset_1},  \eqref{fig:dataset_2}, and \eqref{fig:dataset_3} depict synthetic generative GP datasets used for hyperparameter accuracy evaluation experiments, while Figure~\eqref{fig:dataset_4} and \eqref{fig:dataset_5} show real-world NASA SRTM terrain datasets~\cite{farr2007shuttle} used for assessing prediction performance.}
    \label{fig:datasets_all}
\end{figure*}

We monitor the convergence of each agent using primal $\boldsymbol{r}_i^{(s+1)} = \boldsymbol{\theta}_i^{(s+1)} - \boldsymbol{z}^{(s+1)}$ and the dual residual $\boldsymbol{s}_i^{(s+1)} = \rho_i \left( \boldsymbol{z}^{(s+1)} - \boldsymbol{z}^{(s)}\right)$. These residuals must satisfy the conditions,  $\left\| \boldsymbol{r}_i^{(s+1)} \right\| \le \epsilon_{\text{primal}}$ and $\left\| \boldsymbol{s}_i^{(s+1)} \right\| \le \epsilon_{\text{dual}}$ with tolerances,
\begin{subequations}
\begin{align}\label{eq:convergence_primal}
    \epsilon_{\text{primal}} &= \sqrt{n_{\text{p}}} \epsilon_{\text{abs}} + \epsilon_{\text{rel}} \text{ max} \left\{ \left\| \boldsymbol{\theta}_i^{(s+1)} \right\|, \left\| \boldsymbol{z}^{(s+1)} \right\| \right\}\\\label{eq:convergence_dual}
    \epsilon_{\text{dual}} &= \sqrt{n_{\text{d}}} \epsilon_{\text{abs}} + \epsilon_{\text{rel}} \left\| \rho_i \boldsymbol{u}_i^{(s+1)} \right\|,
\end{align}
\label{eq:convergence_criteria}%
\end{subequations}
where $n_{\text{p}}$ and $n_{\text{d}}$ are the dimensions of the primal $\boldsymbol{\theta}$ and dual $\boldsymbol{u}$ variables, and $\epsilon_{\text{abs}}$, $\epsilon_{\text{rel}}$ are absolute and relative tolerances.


The penalty parameter $\rho_i$ balances the consensus constraint and the local objective in the augmented Lagrangian~\eqref{eq:augmented_lagrangian_equation}. 
Unlike prior works that fix $\rho_i$ to a heuristic value, we adopt a residual-balancing strategy \cite{boyd2011admm} that adjusts $\rho_i$ based on 
primal and dual residuals,
\begin{align}\label{eq:adaptive_rho}
    \rho_i^{(s+1)} =
    \begin{cases}
    \tau_{\text{incr}} \rho_i^{(s)}, & \text{if } \left\| r^{(s)} \right\|  > \beta \left\| s^{(s)} \right\| \\
    \frac{\rho^{(s)}}{\tau_{\text{decr}}}, & \text{if } \left\| s^{(s)} \right\| > \beta \left\| r^{(s)} \right\| \\
    \rho^{(s)}, & \text{otherwise},
\end{cases}
\end{align}
where $\beta$, $\tau_{\text{iccr}}$, $\tau_{\text{decr}} > 1$. In addition, we also adjust the local Lipschitz variable $L_i^{(s+1)}$ using a backtracking line search based on the Armijo condition~\cite{bertsekas1999nonlinear}. In addition, at each iteration, we ensure that the local objective satisfies the condition,
\begin{equation}
\begin{aligned}
    \mathcal{L}_i(\boldsymbol{\theta}_i^{(s+1)}) \le \mathcal{L}_i(\boldsymbol{v}_i) - \frac{c \left\| \nabla_{\boldsymbol{\theta}} \mathcal{L}_i(\boldsymbol{v}_i) \right\|^2}{L_i^{(s+1)} + \rho_i^{(s+1)}},
\end{aligned}
\label{eq:adaptive_lip}
\end{equation}
where $c \in (0, 1)$.~
If the condition fails, $L_i^{(s+1)}$ is reduced by a factor $\tau_{\text{lip}} \in (0, 1)$ until satisfied or a retry limit is reached. 


\newfloat{algorithm}{!t}{lop}
\begin{algorithm}
\caption{dec-pxpGP}
\begin{algorithmic}[1]

\Statex \textbf{Input:} $\mathcal{D}_i = (\boldsymbol{X}_i, \boldsymbol{y}_i)$, $k(\cdot,\cdot)$, $\rho_i$, $L_i$, $\mathcal{N}_i$ , $s_{\textrm{dec-pxpGP}}^{\textrm{end}}$

\Statex \textbf{Output:} $\hat{\boldsymbol{\theta}}$, $\mathcal{D}_{+i}^*$

\For{$i = 1$ to $M$}  \Comment{Sparse Modeling}
    \State $\mathcal{D}_i^*, \boldsymbol{\theta}_{i}^* \leftarrow \texttt{SparseModel}(\mathcal{D}_i) $~\eqref{eq:elbo_final_obj}
    \State $\mathcal{D}_c^* \leftarrow \texttt{Flooding}(\mathcal{D}_i^*, \mathcal{N}_i)$
    \State $\mathcal{D}_{+i}^* = \mathcal{D}_{i} \cup  \mathcal{D}_{c}^*$ \Comment{Local Augmented Dataset}
    \State Initialize $\boldsymbol{\theta}_i^{(1)} = \boldsymbol{\theta}_{i}^*$ \Comment{Warm Start}
\EndFor

\For{$s = 1$ to $s_{\textrm{dec-pxpGP}}^{\textrm{end}}$} \Comment{dec-ADMM optimization}
\ForEach{$i \in \mathcal{V}$}  
    \State Communicate $\boldsymbol{\theta}_i^{(s)}$ with $\mathcal{N}_i$
    \State $\boldsymbol{\alpha}_i^{(s+1)} \leftarrow \texttt{dual}(\boldsymbol{\alpha}_i^{(s)},  \boldsymbol{\theta}_i^{(s)}, \boldsymbol{\theta}_j^{(s)})$ (\ref{eq:dec_pxpgp_alpha_update})
    \State $\boldsymbol{\theta}_i^{(s+1)} \leftarrow  \texttt{primal}(\boldsymbol{\alpha}_i^{(s+1)}, \boldsymbol{\theta}_i^{(s)}, \boldsymbol{\theta}_j^{(s)} \mathcal{D}_{+i}^*)$ 
    (\ref{eq:dec_pxpgp_theta_update})
    \State Update $\rho_i^{(s+1)}$ (\ref{eq:adaptive_rho}), $L_i^{(s+1)}$ (\ref{eq:adaptive_lip})
\EndFor
\EndFor

\State \Return $\hat{\boldsymbol{\theta}}$
\end{algorithmic}
\label{algo:dec_pxpgp}
\end{algorithm}

The details of pxpGP implementation are presented in 
Algorithm~\ref{algo:cent_pxpgp}. First, each agent computes locally a compact sparse GP model by solving ~\eqref{eq:elbo_final_obj} and communicates the local compact pseudo-dataset $\mathcal{D}_i^*$ to the central node. Next, the central node aggregates and broadcasts the communication dataset $\mathcal{D}_c^*$ to all agents. Then, each agent forms the local pseudo-augmented dataset $\mathcal{D}_{+i}^*$ and initializes the local hyperparameter vector based on the local compact sparse GP model $\boldsymbol{\theta}_{i}^*$ to warm-start the optimization. The ADMM optimization begins by communicating $\boldsymbol{\theta}_i^{(s)}$ to the central node to update the primal variable $\boldsymbol{z}^{(s+1)}$~\eqref{eq:pxpgp_z_update} and broadcast the computed value to all agents. Algorithm~\ref{algo:cent_pxpgp} requires coordination with a central node; asynchronous federated GP formulations that rely on different update rules and convergence assumptions are studied in~\cite{shi2025scalable}. Then, each agent computes the primal variable $\boldsymbol{\theta}_i^{(s+1)}$~\eqref{eq:pxpgp_theta_update}, the dual variable $\boldsymbol{u}_i^{(s+1)}$~\eqref{eq:pxpgp_u_update}, while updating the penalty parameter $\rho_i^{(s+1)}$~\eqref{eq:adaptive_rho} and Lipschitz parameter~$L_i^{(s+1)}$~\eqref{eq:adaptive_lip}. The optimization iterates until the primal~\eqref{eq:convergence_primal} and dual residual~\eqref{eq:convergence_dual} converge.

\subsection{Decentralized pxpGP training}


We extend the pxpGP framework to decentralized network topologies by addressing the optimization problem~\eqref{eq:dec_admm_nll_min_problem}. In decentralized pxpGP (dec-pxpGP), each agent $i$ independently optimizes its local GP hyperparameters $\boldsymbol{\theta}_i$ using its local pseudo-augmented dataset $\mathcal{D}_{+i}^*$, while communicating only with its immediate neighbors $\mathcal{N}_i$ over a static, connected undirected graph $\mathcal{G}$. To distribute the pseudo-datasets across the network, we adopt a flooding mechanism that ensures all agents receive the shared communication dataset $\mathcal{D}_c^*$. Similarly to the centralized pxpGP, dec-pxpGP leverages locally learned variational hyperparameters $\boldsymbol{\theta}_i^*$ for warm-start initialization of the local hyperparameters $\boldsymbol{\theta}_i^{(1)}=\boldsymbol{\theta}_i^*$ in the global training stage. Each agent optimizes $\boldsymbol{\theta}_i$ while maintaining consensus on shared variables $\boldsymbol{z}_{ij}$ with its neighbors. Through synchronous iterative updates, all agents progressively converge to a consistent global hyperparameter estimate by following~\cite{kontoudis2024scalable} that yields,
\begin{subequations}
\begin{align}\nonumber 
    \boldsymbol{\theta}_i^{(s+1)} &= \frac{1}{L_i^{(s)} + 2\rho_i \left| \mathcal{N}_i \right|} \left( 
        \rho_i^{(s)} \sum_{j\in \mathcal{N}_i}^{} \boldsymbol{\theta}_j^{(s)}
        -  \nabla_{\boldsymbol{\theta}} \mathcal{L}_i\left(\boldsymbol{\theta}_i^{(s)}\right) \right. \\  \label{eq:dec_pxpgp_theta_update}
    & \left. \quad - \boldsymbol{\alpha}_i^{(s)}  + \left( \rho_i^{(s)} \left| \mathcal{N}_i \right| + L_i^{(s)} \right) \boldsymbol{\theta}_i^{(s)} 
    \right) \\ \label{eq:dec_pxpgp_alpha_update}
    \boldsymbol{\alpha}_i^{(s+1)} &= \boldsymbol{\alpha}_i^{(s)} + \rho_i^{(s)} \left( \left| \mathcal{N}_i \right| \boldsymbol{\theta}_i^{(s+1)} - \sum_{j\in \mathcal{N}_i}^{} \boldsymbol{\theta}_j^{(s+1)} \right), 
\end{align}
\label{eq:dec_pxpgp_iterative_scheme}%
\end{subequations}
where $\left| \mathcal{N}_i \right|$ denotes the number of neighbors of agent $i$ (cardinality) and $\boldsymbol{\alpha}_i$ represents the dual variable. This formulation enables parallel, neighbor-only communication updates, resulting in a scalable and robust approach across varying network topologies.


The dec-pxpGP employs the adaptive residual-balancing strategy for $\rho_i$~\eqref{eq:adaptive_rho} and tunes the Lipschitz parameter $L_i$~\eqref{eq:adaptive_lip} using the Armijo condition (Algorithm~\ref{algo:dec_pxpgp}).

\begin{figure*}[!t]
  \centering
  \includegraphics[width=1.0\textwidth]{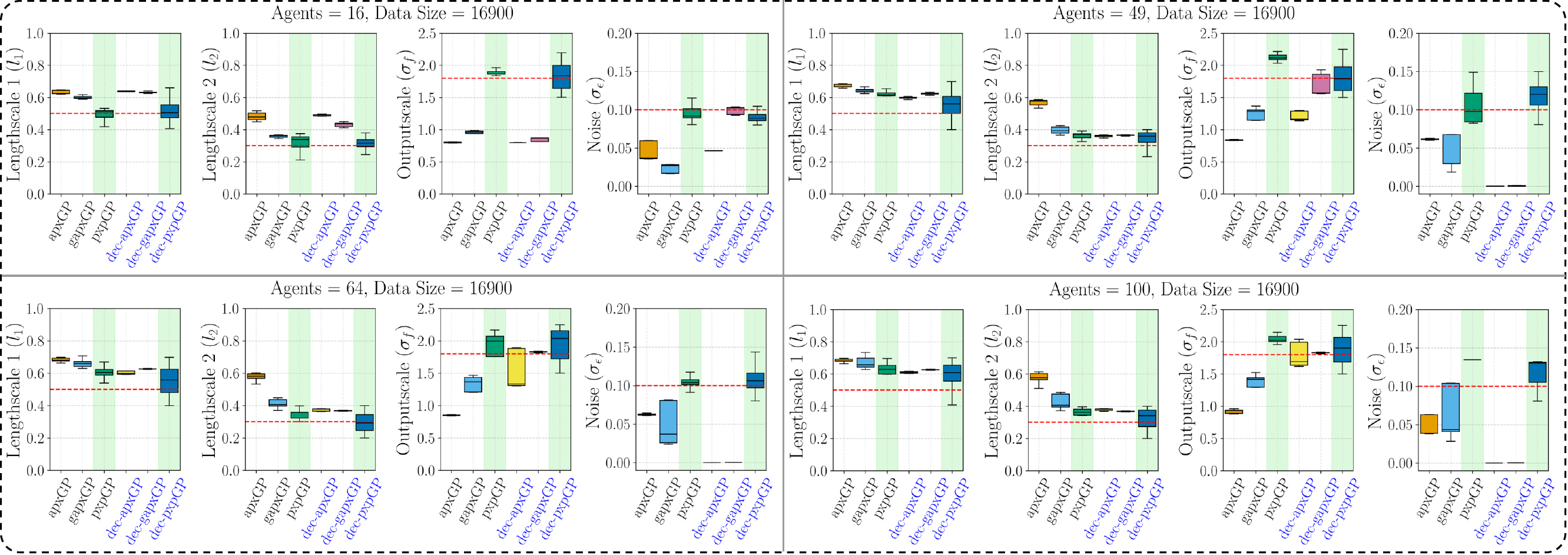}
    \vspace{-15pt}
  \caption{Hyperparameter estimation accuracy of baseline GP methods and proposed pxpGP (highlighted with green background) for centralized (black) and decentralized (blue) setups across fleet sizes $M = \left\{ 16, 49, 64, 100 \right\}$ on a dataset with $N = 16{,}900$. Red dashed lines indicate ground-truth hyperparameters. 
  \label{fig:result_16k}}
  \vspace{-4pt}
\end{figure*}

\section{Numerical Experiments and Results}\label{sec:numericalExperiments}


\begin{figure*}[!t]
  \centering
  \includegraphics[width=1.0\textwidth]{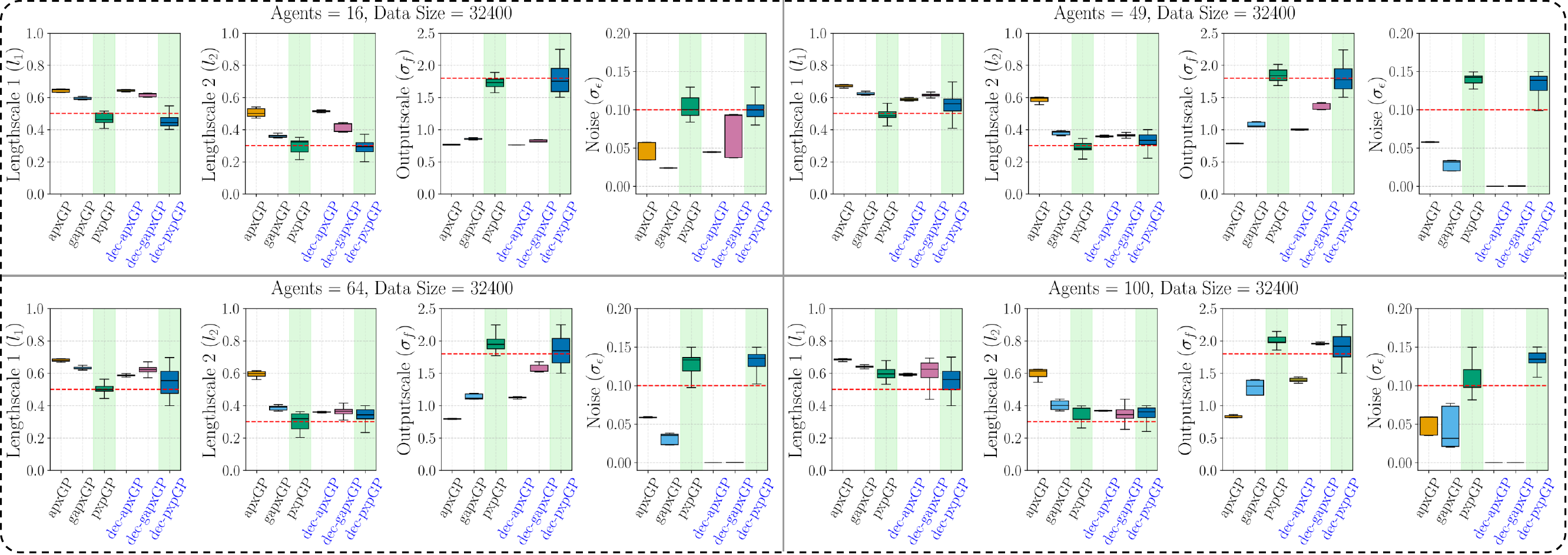}
    \vspace{-15pt}
  \caption{Hyperparameter estimation accuracy of baseline GP methods and proposed pxpGP (highlighted with green background) for centralized (black) and decentralized (blue) setups across fleet sizes $M = \left\{ 16, 49, 64, 100 \right\}$ on a dataset with $N = 32{,}400$. Red dashed lines indicate ground-truth hyperparameters.   
  \label{fig:result_32k}}
  \vspace{-4pt}
\end{figure*}








\begin{table*}[t]
\centering
\caption{Prediction accuracy of the proposed pxpGP and dec-pxpGP frameworks across fleet size $M = {16, 49, 64, 100}$, using a training dataset of size $N = 30,000$ equally distributed among agents and a test dataset size $N_{test} = 300$ per agent, compared with the baseline models (gapxGP and dec-gapxGP~\cite{kontoudis2024scalable}) on the SRTM dataset~\cite{farr2007shuttle}.}
\label{tab:srtm_result}
\renewcommand{\arraystretch}{1}
\small{\begin{tabular}{cccccccccc}
\toprule

& & \multicolumn{4}{c}{Centralized GPs} & \multicolumn{4}{c}{Decentralized GPs}  \\
\cmidrule(lr){3-6} \cmidrule(lr){7-10}

\multirow{2}{*}{Dataset} & \multirow{2}{*}{M} & \multicolumn{2}{c}{\textbf{pxpGP}} & \multicolumn{2}{c}{gapxGP~\cite{kontoudis2024scalable}} & \multicolumn{2}{c}{\textbf{dec-pxpGP}} & \multicolumn{2}{c}{dec-gapxGP~\cite{kontoudis2024scalable}} \\

\cmidrule(lr){3-4} \cmidrule(lr){5-6} \cmidrule(lr){7-8} \cmidrule(lr){9-10}
& & NRMSE $\downarrow$ & NLPD $\downarrow$ & NRMSE $\downarrow$ & NLPD $\downarrow$ & NRMSE $\downarrow$ & NLPD $\downarrow$ & NRMSE $\downarrow$ & NLPD $\downarrow$ \\

\midrule
\multirow{5}{*}{N39W106} 
                         & 16 & \textbf{0.058 $\pm$ 0.012} & \textbf{0.265 $\pm$ 0.057} & $0.071 \pm 0.001 $ & $0.437 \pm 0.009$ & \textbf{0.067 $\pm$ 0.001} & \textbf{0.305 $\pm$ 0.009} & $0.072 \pm 0.007$ & $0.311 \pm 0.093$ \\                         
                         & 49 & $0.080 \pm 0.002$ & \textbf{0.414 $\pm$ 0.033} & \textbf{0.076 $\pm$ 0.002} & $0.487 \pm 0.014$ & \textbf{0.061 $\pm$ 0.002} & \textbf{0.153 $\pm$ 0.038} & $0.062 \pm 0.003$ & $0.213 \pm 0.076$ \\
                         & 64 & $0.081 \pm 0.003$ & \textbf{0.422 $\pm$ 0.037} & \textbf{0.076 $\pm$ 0.002} & $0.492 \pm 0.018$ & \textbf{0.062 $\pm$ 0.002} & \textbf{0.170 $\pm$ 0.029} & $0.062 \pm 0.003$ & $0.204 \pm 0.056$ \\                         
                         & 100 & $0.086 \pm 0.003$ & $0.562 \pm 0.021$ & \textbf{0.081 $\pm$ 0.003} & \textbf{0.521 $\pm$ 0.019} & \textbf{0.067 $\pm$ 0.008} & \textbf{0.219 $\pm$ 0.084} & $0.068 \pm 0.006$ & $0.272 \pm 0.139$ \\
\midrule
\multirow{5}{*}{N37W120} 
                         & 16 & \textbf{0.067 $\pm$ 0.012} & $0.387 \pm 0.130$ & $0.069 \pm 0.002$ & \textbf{0.381 $\pm$ 0.057} & \textbf{0.067 $\pm$ 0.005} & \textbf{0.202 $\pm$ 0.029} & $0.071 \pm 0.001$ & $0.235 \pm 0.070$ \\
                         & 49 & $0.076 \pm 0.004$ & \textbf{0.281 $\pm$ 0.135} & \textbf{0.074 $\pm$ 0.003} & $0.421 \pm 0.067$ & $0.062 \pm 0.002$ & \textbf{0.061 $\pm$ 0.101} & \textbf{0.061 $\pm$ 0.003} & $0.151 \pm 0.114$ \\
                         & 64 & $0.077 \pm 0.004$ & \textbf{0.292 $\pm$ 0.133} & \textbf{0.075 $\pm$ 0.003} & $0.432 \pm 0.062$ & $0.064 \pm 0.004$ & \textbf{0.084 $\pm$ 0.094} & \textbf{0.063 $\pm$ 0.003} & $0.133 \pm 0.093$ \\
                         & 100 & $0.083 \pm 0.004$ & \textbf{0.392 $\pm$ 0.040} & \textbf{0.078 $\pm$ 0.003} & $0.457 \pm 0.066$ & \textbf{0.066 $\pm$ 0.005} & \textbf{0.106 $\pm$ 0.129} & $0.068 \pm 0.006$ & $0.204 \pm 0.123$\\
\midrule
\multirow{5}{*}{N43W080} 
                         & 16 & \textbf{0.057 $\pm$ 0.016} & $0.327 \pm 0.139$ & $0.065 \pm 0.005$ & \textbf{0.314 $\pm$ 0.105} & $0.065 \pm 0.008$ & $0.123 \pm 0.055$ & \textbf{0.062 $\pm$ 0.005} & \textbf{0.161 $\pm$ 0.091} \\
                         & 49 & \textbf{0.071 $\pm$ 0.005} & \textbf{0.193 $\pm$ 0.169} & $0.071 \pm 0.008$ & $0.371 \pm 0.095$ & \textbf{0.054 $\pm$ 0.008} & \textbf{-0.086 $\pm$ 0.178} & $0.061 \pm 0.004$ & $0.091 \pm 0.131$ \\
                         & 64 & \textbf{0.072 $\pm$ 0.004} & \textbf{0.206 $\pm$ 0.165} & $0.072 \pm 0.009$ & $0.397 \pm 0.077$ & $0.060 \pm 0.006$ & \textbf{0.005 $\pm$ 0.130} & \textbf{0.057 $\pm$ 0.007} & $0.067 \pm 0.124$ \\
                         & 100 & $0.082 \pm 0.004$ & \textbf{0.343 $\pm$ 0.035} & \textbf{0.075 $\pm$ 0.005} & $0.432 \pm 0.065$ & \textbf{0.060 $\pm$ 0.008} & \textbf{0.005 $\pm$ 0.182} & $0.067 \pm 0.006$ & $0.178 \pm 0.128$ \\
\bottomrule
\end{tabular}}
\end{table*}

\begin{table*}[t]
\centering
\caption{Computational and Communication Complexity Comparison of Distributed GP Methods}
\label{tab:complexity}
\renewcommand{\arraystretch}{1.3}
\begin{tabular}{lccccc}
\toprule
 & {apxGP \cite{xie2019distributed}} & {gapxGP \cite{kontoudis2024scalable}} & \textbf{pxpGP (Ours)} & {dec-gapxGP}~\cite{kontoudis2024scalable} & \textbf{dec-pxpGP (Ours)} \\  \cmidrule(lr){2-6}

{Time} & $\mathcal{O}\left(\frac{N^3}{M^3}\right)$ & $\mathcal{O}\left(8 \frac{N^3}{M^3}\right)$ & $\mathcal{O}\left(8 \frac{N^3}{M^3}\right) + \mathcal{O}\left(B P^2 + P^3\right)$ & $\mathcal{O}\left(8 \frac{N^3}{M^3}\right)$ & $\mathcal{O}\left(8 \frac{N^3}{M^3}\right) + \mathcal{O}\left(B P^2 + P^3\right)$ \\ 
\cmidrule(lr){2-6}

{Space} & $\mathcal{O}(\xi)$ & $\mathcal{O}\left(2\xi + \frac{2N^2}{M^2}\right)$ & $\mathcal{O}\left(2\xi + \frac{2N^2}{M^2} + P^2\right)$ & $\mathcal{O}\left(2\xi + \frac{2N^2}{M^2}\right)$ & $\mathcal{O}\left(2\xi + \frac{2N^2}{M^2} + P^2\right)$ \\ 
\cmidrule(lr){2-6}

{Comms} & $\mathcal{O}\left(s^{\text{end}} M (D + 2)\right)$ & $\mathcal{O}\left(s^{\text{end}} M (D + 2) \right)$ & $\mathcal{O}\left(s^{\text{end}} M (D + 2) \right)$ & $\mathcal{O}\left(s^{\text{end}} M(D + 2)\right)$ & $\mathcal{O}\left(s^{\text{end}} M (D + 2)\right)$ \\
\bottomrule
\end{tabular}
\end{table*}

To illustrate the efficacy of the proposed \textbf{pxpGP} and \textbf{dec-pxpGP} training method, we conduct numerical experiments with both synthetic and real-world datasets, and compare against existing distributed GP methods~\cite{xie2019distributed},~\cite{kontoudis2024scalable}. 
For our experiments, we generate 2D synthetic datasets using generative GP functions with known hyperparameters $\boldsymbol{\theta} = (l_1, l_2, \sigma_f, \sigma_{\epsilon})^{\intercal} = (0.7, 0.5, 1.8, 0.1)^{\intercal}$ for controlled benchmarking of GP hyperparameters, and used the NASA Shuttle Radar Topography Mission (SRTM) terrain elevation dataset~\cite{farr2007shuttle} to evaluate the prediction performance.

We perform synthetic dataset experiments with two different dataset sizes, $N = 16,900$ and $N=34,900$. For the real-world SRTM dataset, we use 3 tiles (N39W106, N37W120, N43W080), each with $N=30,000$ training samples divided among agents and assign $N_{\text{test}} = 300$ test samples to each  agent. Each training dataset is spatially and sequentially partitioned into equal-sized local datasets, satisfying Assumption~\ref{assumption:dist_dataset}, across varying fleet sizes $M \in \left\{ 16, 49, 64, 100 \right\}$. 
For real-world dataset experiments, we use a single consistent global test dataset $N_{\text{test}}$ among all agents to evaluate prediction performance. In all experiments, the number of inducing points per agent is chosen as $P = \max\{(N_{i}/M), 4\}$, where $N_{i}$ is the dataset size of agent $i$.
All experiments are implemented in Python using PyTorch and GPyTorch~\cite{gardner2018gpytorch}, running on a workstation with an Intel Core i7-14700 CPU, 62 GB RAM, and an NVIDIA GeForce RTX 4080 GPU with 16 GB VRAM.

The proposed pxpGP and dec-pxpGP frameworks are evaluated by benchmarking their hyperparameter estimation accuracy against several baseline methods, including the global full GP, centralized variants (apxGP~\cite{xie2019distributed} and gapxGP~\cite{kontoudis2024scalable}), and their decentralized counterparts (dec-apxGP~\cite{kontoudis2024scalable} and dec-gapxGP~\cite{kontoudis2024scalable}). The predictive performance of proposed pxpGP and dec-pxpGP is compared against gapxGP and dec-gapxGP. The apxGP method is excluded from this comparison, since the gapxGP formulation provides a superior and more scalable alternative.
For the centralized methods, the penalty and Lipschitz parameters are fixed at $\rho = 5$, $L_i = 10$ respectively, with convergence tolerances set to $\epsilon_{\text{abs}}=10^{-5}$ and a maximum of $1,000$ ADMM iterations.
In contrast, pxpGP uses adaptive updates of $\rho$ and $L_i$, initialized with $\rho^{(1)} =1.0 $ and $L_i^{(1)} = 5.0$, with the outer ADMM iterations are capped at $s^{\text{end}} = 500$. 
For the decentralized experiments, we adopt a minimal connected graph topology in which each agent has a maximum neighborhood degree of $\left| \mathcal{N} \right| = 2$, yielding a 
$1$-connected network that satisfies the standard connectivity assumption for consensus while providing a lower-bound scenario on communication redundancy and mixing speed.
This graph topology and data distribution represent the worst-case connectivity and worst-case data allocation conditions. As network connectivity increases or local regions begin to overlap, all methods demonstrate improved performance.


The evaluation focuses on three aspects: i) \textbf{hyperparameter estimation accuracy} 
relative to the ground-truth using the synthetic dataset; ii) \textbf{prediction performance} 
relative to the ground-truth on the real-world SRTM dataset; and iii) \textbf{computational and communication complexity} 
compared to baseline methods.

\subsection{Hyperparameter Accuracy Estimate}
The accuracy of the hyperparameter estimation 
is a key indicator of model consistency and scalability across 
agents. For the smaller synthetic dataset ($N = 16,900$), Fig.~\ref{fig:result_16k}, pxpGP and dec-pxpGP remain close to the ground-truth hyperparameters for all fleet sizes $M$, while the accuracy of baseline methods degrades noticeably as the number of agents $M$ increases.~
For the larger synthetic dataset ($N = 34,900$), Fig.~\ref{fig:result_32k}, all methods benefit from the increased data volume, but the pxpGP and dec-pxpGP still provide the most accurate and stable estimates across all fleet sizes, particularly in larger networks.


\subsection{Prediction Performance}
Reliable prediction and uncertainty estimation are key to evaluating model performance in distributed multi-robot learning. To evaluate these parameters, we assess the predictive performance of the proposed pxpGP and dec-pxpGP frameworks on three tiles of the real-world SRTM terrain dataset, comparing them with baseline gapxGP and dec-gapxGP methods. As summarized in Table~\ref{tab:srtm_result}, both pxpGP and dec-pxpGP achieve comparable or nearly identical Normalized Root Mean Square Error (NRMSE) values relative to their respective baselines.

More importantly, the proposed methods consistently yield substantially lower Negative Log Predictive Density (NLPD) values, which measure predictive uncertainty. Lower NLPD corresponds to more accurate and confident predictions, particularly for larger fleet sizes and 
non-stationary datasets such as tiles N37W120 and N43W080.
Thus, the reduced NLPD values demonstrate that the proposed methods provide superior uncertainty quantification and higher model confidence compared to baseline approaches. Overall, both pxpGP and dec-pxpGP maintain stable prediction accuracy and well-calibrated uncertainty across diverse datasets and large-scale networks.



\subsection{Computational Analysis}
We compare the computational efficiency and scalability of all distributed GP training methods in Table~\ref{tab:complexity}, which summarizes their time, space, and communication complexity, where $\xi = N^2/M^2+D(N/M)$. While pxpGP and dec-pxpGP 
add a one-time computational overhead from local sparse GP training ($\mathcal{O}\left(B P^2 + P^3\right)$) to generate the local compact pseudo-dataset $\mathcal{D}_i^*$, where $B$ is the batch size. This overhead is offset 
by: i) improved initialization, which requires fewer ADMM iterations to converge;
and ii) adaptive tuning of $\rho_i, \text{ and } L_i$, which minimizes manual adjustments and accelerates convergence. 


Unlike gapxGP and dec-gapxGP~\cite{kontoudis2024scalable}, which rely on random sampling from local datasets, the proposed pxpGP and dec-pxpGP share compact pseudo-representations that 
enhance data privacy, and reduce communication overhead. Moreover, the dataset heterogeneity in gapxGP often leads to ill-conditioned covariance matrices, resulting in numerical instability during inversion, especially for large-scale networks. In contrast, pxpGP maintains numerical robustness through the use of optimized sparse datasets, warm-start initialization, and adaptive parameter selection, all of which contribute to faster convergence. 
Additionally, the proposed pxpGP employs warm-start initialization with near-optimal hyperparameters, reducing the need for strict convergence tolerances and requiring fewer ADMM iterations. The latter leads to improved hyperparameter estimation accuracy for large-scale networks, while lowering computational and communication costs.

\section{Conclusion 
}\label{sec:conclusion}

In this work, we introduced \textbf{pxpGP} and \textbf{dec-pxpGP}, scalable and federated GP methods for large-scale multi-robot learning. By combining sparse variational inference with boundary and repulsive penalty terms, pxpGP constructs informative and well-distributed shared pseudo-datasets that enhance global data representation. 
The proposed centralized and decentralized variants demonstrate efficient and accurate hyperparameter estimation, and superior predictive performance in numerical experiments, spanning fleet sizes from $16$ to $100$ agents, while preserving data privacy. 

These results underscore the scalability, efficiency, and robustness of the proposed method, positioning pxpGP as a strong candidate for real-world federated learning applications in large-scale multi-robot systems. Beyond scalable model learning, the proposed framework establishes a foundation for model-based control and decision-making, maintaining consistent uncertainty estimates under limited communication. This makes it particularly suited for 
cooperative exploration, where reliable modeling directly influences control and coordination strategies.

\balance
\bibliographystyle{ACM-Reference-Format} 
\bibliography{sample}

@article{kontoudis2025multi,
  title={Multi-Agent Federated Learning Using Covariance-Based Nearest Neighbor {Gaussian} Processes},
  author={Kontoudis, George P and Stilwell, Daniel J},
  journal={IEEE Transactions on Machine Learning in Communications and Networking},
  volume={4},
  pages={115--138},
  year={2025},
  publisher={IEEE}
}

@article{shi2025scalable,
  title={Scalable Asynchronous Federated Modeling for Spatial Data},
  author={Shi, Jianwei and Abdulah, Sameh and Sun, Ying and Genton, Marc G},
  journal={arXiv preprint arXiv:2510.01771},
  year={2025}
}

@inproceedings{llorente2025decentralized,
  title={Decentralized Online Ensembles of {G}aussian Processes for Multi-Agent Systems},
  author={Llorente, Fernando and Waxman, Daniel and Djuri{\'c}, Petar M},
  booktitle={IEEE International Conference on Acoustics, Speech and Signal Processing},
  pages={1--5},
  year={2025}
}

@article{xie2019distributed,
  title={Distributed {G}aussian processes hyperparameter optimization for big data using proximal {ADMM}},
  author={Xie, Ang and Yin, Feng and Xu, Yue and Ai, Bo and Chen, Tianshi and Cui, Shuguang},
  journal={IEEE Signal Processing Letters},
  volume={26},
  number={8},
  pages={1197--1201},
  year={2019},
  publisher={IEEE}
}

@article{snelson2005sparse,
  title={Sparse {G}aussian processes using pseudo-inputs},
  author={Snelson, Edward and Ghahramani, Zoubin},
  journal={Advances in Neural Information Processing Systems},
  volume={18},
  year={2005}
}

@article{xu2019wireless,
  title={Wireless traffic prediction with scalable {G}aussian process: Framework, algorithms, and verification},
  author={Xu, Yue and Yin, Feng and Xu, Wenjun and Lin, Jiaru and Cui, Shuguang},
  journal={IEEE Journal on Selected Areas in Communication},
  volume={37},
  number={6},
  pages={1291--1306},
  year={2019},
  publisher={IEEE}
}

@article{hong2016convergence,
  title={Convergence analysis of alternating direction method of multipliers for a family of nonconvex problems},
  author={Hong, Mingyi and Luo, Zhi-Quan and Razaviyayn, Meisam},
  journal={SIAM Journal on Optimization},
  volume={26},
  number={1},
  pages={337--364},
  year={2016},
  publisher={SIAM}
}

@article{shi2014linear,
  title={On the linear convergence of the {ADMM} in decentralized consensus optimization},
  author={Shi, Wei and Ling, Qing and Yuan, Kun and Wu, Gang and Yin, Wotao},
  journal={IEEE Transactions on Signal Processing},
  volume={62},
  number={7},
  pages={1750--1761},
  year={2014},
  publisher={IEEE}
}

@article{queralta2020collaborative,
  title={Collaborative multi-robot search and rescue: Planning, coordination, perception, and active vision},
  author={Queralta, Jorge Pena and Taipalmaa, Jussi and Pullinen, Bilge Can and Sarker, Victor Kathan and Gia, Tuan Nguyen and Tenhunen, Hannu and Gabbouj, Moncef and Raitoharju, Jenni and Westerlund, Tomi},
  journal={IEEE Access},
  volume={8},
  pages={191617--191643},
  year={2020},
  publisher={IEEE}
}

@book{rasmussen2006gaussian,
  title={Gaussian Processes for Machine Learning},
  author={Rasmussen, Carl Edward and Williams, Christopher KI},
  edition = {2},
  year={2006},
  publisher={Cambridge, MA, USA: MIT Press}
}

@article{jang2020multi,
  title={Multi-Robot Active Sensing and Environmental Model Learning With Distributed {G}aussian Process},
  author={Jang, Dohyun and Yoo, Jaehyun and Son, Clark Youngdong and Kim, Dabin and Kim, H Jin},
  journal={IEEE Robotics and Automation Letters},
  volume={5},
  number={4},
  pages={5905--5912},
  year={2020},
  publisher={IEEE}
}

@inproceedings{stump2011multi,
  title={Multi-robot persistent surveillance planning as a vehicle routing problem},
  author={Stump, Ethan and Michael, Nathan},
  booktitle={IEEE International Conference on Automation Science and Engineering},
  pages={569--575},
  year={2011}
}

@inproceedings{titsias2009variational,
  title={Variational learning of inducing variables in sparse Gaussian processes},
  author={Titsias, Michalis},
  booktitle={Artificial intelligence and statistics},
  pages={567--574},
  year={2009},
  organization={PMLR}
}

@mastersthesis{norton2022efficient,
  title={Efficient and Adaptive Decentralized Sparse Gaussian Process Regression for Environmental Sampling Using Autonomous Vehicles},
  author={Norton, Tanner A},
  year={2022},
  school={Brigham Young University}
}

@article{farr2007shuttle,
  title={The shuttle radar topography mission},
  author={Farr, Tom G and Rosen, Paul A and Caro, Edward and Crippen, Robert and Duren, Riley and Hensley, Scott and Kobrick, Michael and Paller, Mimi and Rodriguez, Ernesto and Roth, Ladislav and others},
  journal={Reviews of geophysics},
  volume={45},
  number={2},
  year={2007},
  publisher={Wiley Online Library}
}

@book{boyd2011admm,
year = {2011},
volume = {3},
journal = {Foundations and Trends in Machine Learning},
title = {Distributed Optimization and Statistical Learning via the Alternating Direction Method of Multipliers},
number = {1},
author = {Stephen Boyd and Neal Parikh and Eric Chu and Borja Peleato and Jonathan Eckstein}
}

@book{gramacy2020surrogates,
  title = {Surrogates: {G}aussian Process Modeling, Design and Optimization for the Applied Sciences},
  author = {Robert B. Gramacy},
  publisher = {Chapman Hall/CRC},
  address = {Boca Raton, Florida},
  year = {2020}
}

@book{bertsekas1999nonlinear,
  title={Nonlinear programming},
  author={Bertsekas, Dimitri P},
  year={1999},
  publisher={Athena Scientific}
}

@article{das2015data,
  title={Data-driven robotic sampling for marine ecosystem monitoring},
  author={Das, Jnaneshwar and Py, Fr{\'e}d{\'e}ric and Harvey, Julio BJ and Ryan, John P and Gellene, Alyssa and Graham, Rishi and Caron, David A and Rajan, Kanna and Sukhatme, Gaurav S},
  journal={The International Journal of Robotics Research},
  volume={34},
  number={12},
  pages={1435--1452},
  year={2015},
  publisher={SAGE Publications Sage UK: London, England}
}

@article{gardner2018gpytorch,
  title={{GP}y{T}orch: Blackbox matrix-matrix {G}aussian process inference with {GPU} acceleration},
  author={Gardner, Jacob and Pleiss, Geoff and Weinberger, Kilian Q and Bindel, David and Wilson, Andrew G},
  journal={Advances in Neural Information Processing Systems},
  volume={31},
  year={2018}
}

@article{gielis2022critical,
  title={A critical review of communications in multi-robot systems},
  author={Gielis, Jennifer and Shankar, Ajay and Prorok, Amanda},
  journal={Current Robotics Reports},
  volume={3},
  number={4},
  pages={213--225},
  year={2022},
  publisher={Springer}
}

@article{ghaffari2018gaussian,
  title={Gaussian processes autonomous mapping and exploration for range-sensing mobile robots},
  author={Ghaffari Jadidi, Maani and Valls Miro, Jaime and Dissanayake, Gamini},
  journal={Autonomous Robots},
  volume={42},
  pages={273--290},
  year={2018},
  publisher={Springer}
}

@article{suryan2020learning,
  title={Learning a Spatial Field in Minimum Time with a Team of Robots},
  author={Suryan, Varun and Tokekar, Pratap},
  journal={IEEE Transactions on Robotics},
  volume={36},
  number={5},
  pages={1562--1576},
  year={2020},
  publisher={IEEE}
}

@article{liu2020gaussian,
  title={When {G}aussian process meets big data: A review of scalable {GP}s},
  author={Liu, Haitao and Ong, Yew-Soon and Shen, Xiaobo and Cai, Jianfei},
  journal={IEEE Transactions on Neural Networks and Learning Systems},
  year={2020},
  volume={31},
  number={11},
  pages={4405--4423},
  publisher={IEEE}
}

@article{brancato2024adaptive,
  title={Adaptive sampling of a stationary {G}aussian spatial process by a team of robots with heterogeneous dynamics and measurement noise variance},
  author={Brancato, Michael and Wolek, Artur},
  journal={IEEE Access},
  volume={12},
  pages={94407-94423},
  year={2024},
  publisher={IEEE}
}

@article{green2024distributed,
  title={Distributed {G}aussian Processes with Uncertain Inputs},
  author={Green, Peter L},
  journal={IEEE Access},
  volume={12},
  pages={176087-176093},
  year={2024},
  publisher={IEEE}
}

@article{norton2023decentralized,
  title={Decentralized sparse {G}aussian process regression with event-triggered adaptive inducing points},
  author={Norton, Tanner and Stagg, Grant and Ward, Derek and Peterson, Cameron K},
  journal={Journal of Intelligent \& Robotic Systems},
  volume={108},
  number={4},
  pages={72},
  year={2023},
  publisher={Springer}
}

@article{yue2024federated,
  title={Federated {G}aussian Process: Convergence, Automatic Personalization and Multi-fidelity Modeling},
  author={Yue, Xubo and Kontar, Raed},
  journal={IEEE Transactions on Pattern Analysis and Machine Intelligence},
  year={2024},
  volume={46},
  number={6},
  pages={4246-4261},
  publisher={IEEE}
}

@article{corah2019communication,
  title={Communication-efficient planning and mapping for multi-robot exploration in large environments},
  author={Corah, Micah and O’Meadhra, Cormac and Goel, Kshitij and Michael, Nathan},
  journal={IEEE Robotics and Automation Letters},
  volume={4},
  number={2},
  pages={1715--1721},
  year={2019},
  publisher={IEEE}
}

@article{kontoudis2024scalable,
  title={Scalable, Federated {G}aussian Process Training for Decentralized Multi-Agent Systems},
  author={Kontoudis, George P and Stilwell, Daniel J},
  journal={IEEE Access},
  volume={12},
  pages={77800-77815},
  year={2024},
  publisher={IEEE}
}

@inproceedings{masaba2023multi,
  title={Multi-robot adaptive sampling based on mixture of experts approach to modeling non-stationary spatial fields},
  author={Masaba, Kizito and Li, Alberto Quattrini},
  booktitle={IEEE International Symposium on Multi-Robot and Multi-Agent Systems},
  pages={191--198},
  year={2023}
}

@inproceedings{kontoudis2023decentralized,
  title={Decentralized federated learning using {G}aussian processes},
  author={Kontoudis, George P and Stilwell, Daniel J},
  booktitle={IEEE International Symposium on Multi-Robot and Multi-Agent Systems},
  pages={1--7},
  year={2023}
}

@inproceedings{moreno2021modular,
  title={Modular {G}aussian processes for transfer learning},
  author={Moreno-Mu{\~n}oz, Pablo and Art{\'e}s, Antonio and Alvarez, Mauricio},
  booktitle={Advances in Neural Information Processing Systems},
  volume={34},
  pages={24730--24740},
  year={2021}
}

@inproceedings{santos2021multi,
  title={Multi-robot learning and coverage of unknown spatial fields},
  author={Santos, Maria and Madhushani, Udari and Benevento, Alessia and Leonard, Naomi Ehrich},
  booktitle={IEEE International Symposium on Multi-Robot and Multi-Agent Systems},
  pages={137--145},
  year={2021}
}

@inproceedings{luo2018adaptive,
  title={Adaptive sampling and online learning in multi-robot sensor coverage with mixture of {G}aussian processes},
  author={Luo, Wenhao and Sycara, Katia},
  booktitle={IEEE International Conference on Robotics and Automation},
  pages={6359--6364},
  year={2018}
}

@inproceedings{kontoudis2021decentralized,
  title={Decentralized nested {G}aussian processes for multi-robot systems},
  booktitle = {IEEE International Conference on Robotics and Automation},
  author={Kontoudis, George P and Stilwell, Daniel J},
  pages = {8881--8887},
  year = {2021}
}

@inproceedings{hoang2019collective,
  title={Collective online learning of {G}aussian processes in massive multi-agent systems},
  author={Hoang, Trong Nghia and Hoang, Quang Minh and Low, Kian Hsiang and How, Jonathan},
  booktitle={AAAI Conference on Artificial Intelligence},
  volume={33},
  number={01},
  pages={7850--7857},
  year={2019}
}

@inproceedings{deisenroth2015distributed,
  title={Distributed {G}aussian processes},
  author={Deisenroth, Marc and Ng, Jun Wei},
  booktitle={International Conference on Machine Learning},
  pages={1481--1490},
  year={2015}
}


\end{document}